\documentclass[10pt, journal, letterpaper]{IEEEtran}
\usepackage{amsmath,amsfonts}
\usepackage{array}
\usepackage[caption=false]{subfig}
\usepackage{textcomp}
\usepackage{stfloats}
\usepackage{url}
\usepackage{verbatim}
\usepackage{graphicx}

\usepackage[style=ieee, doi=false, url=false, isbn=false, giveninits=true, backend=biber, sorting=none, sortcites=true, date=year]{biblatex}
\DeclareSourcemap{
  \maps[datatype=bibtex]{
    \map{
      \step[fieldset=primaryClass, null]
      \step[fieldset=publisher, null]
      \step[fieldset=editor, null]
    }
  }
}
\usepackage{xcolor}
\definecolor{linkcolor}{rgb}{0.0, 0.0, 0.0}  
\definecolor{citecolor}{rgb}{0.0, 0.6, 0.3}  
\definecolor{urlcolor}{rgb}{0.6, 0.0, 0.3}   

\usepackage[colorlinks=true, linkcolor=linkcolor, citecolor=citecolor, urlcolor=urlcolor]{hyperref}

\usepackage{url}

\usepackage{booktabs}
\usepackage{multirow}
\usepackage{makecell}

\usepackage{algorithmicx}%
\usepackage{algorithm}
\usepackage{algpseudocode}%

\usepackage{bm}

\usepackage[nameinlink, sort, capitalize]{cleveref}
\Crefname{section}{Section}{Sections}
\Crefname{table}{Table}{Tables}

\usepackage[acronym]{glossaries}

\usepackage{tikz}

\usepackage{lipsum}

\usepackage{multicol}
\usepackage{setspace} 
\usepackage{mdframed} 

\usepackage{wrapfig}
\usepackage{orcidlink}

\definecolor{Honeydew}{RGB}{240, 255, 240}
\definecolor{FT}{RGB}{255, 241, 229}

\newacronym{aep}{AEP}{Asymptotic Equipartition Property}
\newacronym{ai}{AI}{Artificial Intelligence}
\newacronym{atroc}{ATROC}{Aligned Task- and Reconstruction-Oriented Communication}
\newacronym{awgn}{AWGN}{Additive White Gaussian Noise}
\newacronym{ber}{BER}{Bit-Error Rate}
\newacronym{bpp}{BPP}{bits per pixel}
\newacronym{csi}{CSI}{Channel State Information}
\newacronym{cvae}{CVAE}{Conditional Variational Autoencoder}
\newacronym{dnn}{DNN}{Deep Neural Network}
\newacronym{dl}{DL}{Deep Learning}
\newacronym{e2e}{E2E}{End-to-End}
\newacronym{fid}{FID}{Fréchet Inception Distance}
\newacronym{gan}{GAN}{Generative Adversarial Network}
\newacronym{ib}{IB}{Information Bottleneck}
\newacronym{iot}{IoT}{Internet of Things}
\newacronym{jscc}{JSCC}{Joint Source-Channel Coding}
\newacronym{kl}{KL}{Kullback-Leibler}
\newacronym{ldpc}{LDPC}{Low-Density Parity-Check}
\newacronym{mse}{MSE}{Mean Square Error}
\newacronym{msssim}{MS-SSIM}{Multi-Scale Structural Similarity}
\newacronym{psnr}{PSNR}{Peak Signal-to-Noise Ratio}
\newacronym{qam}{QAM}{Quadrature Amplitude Modulation}
\newacronym{rf}{RF}{Radio Frequency}
\newacronym{snr}{SNR}{Signal-to-Noise Ratio}
\newacronym{ssim}{SSIM}{Structural Similarity}
\newacronym{tgcp}{TGCP}{Trajectory-Guided Control Prediction}
\newacronym{tscc}{TSCC}{Task-oriented Source-Channel Coding}
\newacronym{vae}{VAE}{Variational Autoencoder}
\newacronym{vib}{VIB}{Variational Information Bottleneck}
\newacronym{v2x}{V2X}{Vehicle-to-Everything}
\newcommand*{\dif}{\mathop{}\!\mathrm{d}}
\renewcommand{\Re}{\operatorname{Re}}
\renewcommand{\Im}{\operatorname{Im}}

\begin{document}

\title{Aligning Task- and Reconstruction-Oriented Communications for Edge Intelligence}

\author{
Yufeng~Diao,~\IEEEmembership{Graduate~Student~Member,~IEEE},~%
Yichi~Zhang,~\IEEEmembership{Graduate~Student~Member,~IEEE},~%
\\%
Changyang~She,~\IEEEmembership{Senior~Member,~IEEE},~%
Philip~Guodong~Zhao,~\IEEEmembership{Senior~Member,~IEEE},
and~Emma~Liying~Li,~\IEEEmembership{Member,~IEEE}%

\thanks{Yufeng Diao is with the School of Computing Science, University of Glasgow, UK (e-mail: y.diao.1@research.gla.ac.uk).}%
\thanks{Yichi Zhang is with the Department of Computer Science, University of Manchester, UK. Part of this work was done when he was with the James Watt School of Engineering, University of Glasgow, UK (e-mail: yichi.zhang@postgrad.manchester.ac.uk).}%
\thanks{Changyang She is with the School of Electrical and Information Engineering, University of Sydney, Australia (e-mail: shechangyang@gmail.com).}%
\thanks{Philip Guodong Zhao is with the Department of Computer Science, University of Manchester, UK (e-mail: philip.zhao@manchester.ac.uk).}%
\thanks{Emma Liying Li is with the School of Computing Science, University of Glasgow, UK (e-mail: liying.li@glasgow.ac.uk).}%
}

\markboth{This paper has been accepted for publication in IEEE Journal on Selected Areas in Communications (JSAC).}%
{}


\maketitle
\begin{abstract}

Existing communication systems aim to reconstruct the information at the receiver side, and are known as reconstruction-oriented communications. This approach often falls short in meeting the real-time, task-specific demands of modern AI-driven applications such as autonomous driving and semantic segmentation. As a new design principle, task-oriented communications have been developed. However, it typically requires joint optimization of encoder, decoder, and modified inference neural networks, resulting in extensive cross-system redesigns and compatibility issues.
This paper proposes a novel communication framework that aligns reconstruction-oriented and task-oriented communications for edge intelligence. The idea is to extend the Information Bottleneck (IB) theory to optimize data transmission by minimizing task-relevant loss function, while maintaining the structure of the original data by an information reshaper. Such an approach integrates task-oriented communications with reconstruction-oriented communications, where a variational approach is designed to handle the intractability of mutual information in high-dimensional neural network features. We also introduce a joint source-channel coding (JSCC) modulation scheme compatible with classical modulation techniques, enabling the deployment of AI technologies within existing digital infrastructures. The proposed framework is particularly effective in edge-based autonomous driving scenarios. Our evaluation in the Car Learning to Act (CARLA) simulator demonstrates that the proposed framework significantly reduces bits per service by 99.19\% compared to existing methods, such as JPEG, JPEG2000, and BPG, without compromising the effectiveness of task execution.
\end{abstract}
\glsresetall


\section{Introduction}
\label{sec:introduction}
Reconstruction-oriented communications are designed to recover the transmitted information at the receiver sides, often involving traditional source and channel coding techniques. This approach is commonly used in systems where the fidelity of the information is paramount, such as in audio or video streaming services. The structure of separate source and channel coding, a cornerstone in the design of communication systems, has been shown to be theoretically optimal via \gls{aep} with infinitely long source and channel blocks \cite{Cover_1991_EoI}. However, in practical scenarios, this separation often leads to inefficiencies and suboptimal performance, particularly for \gls{ai} driven applications \cite{Bourtsoulatze_2019_DJS}.

The pervasive advancement of AI technologies, particularly in the context of deep learning, presents novel challenges for future communication systems, where the throughput required by AI agents could be much higher than that of human users.

Recent developments in deep learning have shown that \gls{jscc} can potentially address some of these inefficiencies and outperform traditional separate coding designs. This approach is especially potent in environments where traditional methods struggle to keep pace with the data demands of \gls{ai}-driven applications. However, \gls{jscc}-based reconstruction-oriented communications, which focus on accurately reconstructing a signal on receiver sides, often waste communication resources by transmitting task-agnostic information \cite{Kurka_2020_DfD}.

To address these issues, task-oriented communication has emerged as a key technology and has attracted significant research interests \cite{Shao_2022_LTO, Shao_2023_TOC, Stavrou_2022_ARD, Shi_2023_TOC}. Using the capabilities of deep learning, task-oriented \gls{jscc} focuses on transmitting task-specific information, thus improving efficiency and reducing the data rate in critical applications. This requires joint optimization of the \gls{jscc} and inference network, which must be co-designed for effective task-oriented communication \cite{Shao_2022_LTO}. Note that existing \gls{jscc} designs are mainly based on analog communication principles \cite{Bourtsoulatze_2019_DJS} and cannot be integrated with existing digital communication infrastructures.

Furthermore, cloud-based services introduce unacceptable latency for real-time applications, such as autonomous driving \cite{Zhang_2020_MEI, Liu_2019_ECf}. To mitigate this issue, \textit{edge inference} \cite{Li_2018_EIO, Shao_2022_LTO, Jankowski_2021_WIR} has become a promising approach, enabling quick response to real-time AI applications. However, widely deployed AI agents bring significant communication loads to communication systems. Emergent methods based on \gls{jscc} have shown great potential to solve this problem \cite{Jankowski_2021_WIR, Shao_2020_BAE, Jankowski_2020_JDE}.

Recognizing these multifaceted challenges, there is a growing interest in developing communication systems that are not only task-oriented but also aligned with reconstruction-oriented communication frameworks. This has led to the proposition of what we refer to as \gls{atroc}, which aims to bridge the gap between the efficiency of task-specific data transmission and the robustness of reconstruction-oriented communications, enabling the seamless integration of \gls{ai} technologies with existing network infrastructures.
\subsection{Related Works}
\subsubsection{Joint Source-Channel Coding}
The rapid evolution of deep learning has significantly influenced communication system designs, aiming to achieve or even surpass the Shannon limit. Deep learning based \gls{jscc} has emerged as a robust solution in scenarios characterized by limited bandwidth and low \gls{snr}. Research in deep \gls{jscc} for reconstruction-oriented communication \cite{Kurka_2019_SRo, Bourtsoulatze_2019_DJS, Kurka_2020_DfD} has demonstrated its superiority over traditional source coding methods, such as JPEG \cite{Wallace_1992_TJs} and JPEG2000 \cite{Taubman_2002_JIc}, as well as channel coding techniques, such as LDPC codes \cite{Gallager_1962_Ldp}, particularly in environments with low \gls{snr}.

Existing reconstruction-oriented communication research primarily focused on data-centric metrics (e.g., \gls{psnr} \cite{Kurka_2019_SRo, Tung_2022_DQC, Yang_2022_OGD, Tung_2022_DLA, Kurka_2020_DfD}, \gls{ssim} \cite{Bourtsoulatze_2019_DJS, Tung_2022_DLA, Yang_2022_OGD, Kurka_2020_DfD}, and \gls{msssim} \cite{Yang_2022_OGD, Tung_2022_DLA, Kurka_2020_DfD}) to evaluate the effectiveness of deep \gls{jscc}. However, these metrics often lead to suboptimal task performance since high-fidelity reconstructions are not always necessary from the machine's perspective, whereas task-specific semantic information plays the most important role \cite{Chaccour_2024_LDM, Yang_2023_SCf, Qin_2021_ScP, Strinati_2024_GOa, Pandey_2023_GOC, Kang_2023_PSi}. For example, in text transmission, the fidelity of words might be compromised to improve communication efficiency while still conveying the intended meanings \cite{Xie_2021_DLE, Farsad_2018_DLf}. Similarly, in image transmission, image fidelity can be sacrificed for less communication overhead and higher task performance \cite{Hu_2022_RSC, Diao_2024_TOS, Diao_2024_TTG}. 

Nonetheless, existing works, such as \cite{Shao_2023_SCW}, assumed that the amplitudes and phases of channel symbols are analog. Thus, we cannot implement them directly in digital communication systems \cite{_2020_ISf}. To address this issue, the authors of \cite{Choi_2019_NJS} explored image transmission over the discrete channel (binary symmetric channel) using variational learning with a Bernoulli prior. This work was further extended by the authors of \cite{Song_2020_Inj}, who introduced adversarial regularization to enhance robustness. Furthermore, recent works \cite{Tung_2022_DQC, Tung_2022_DQCa} investigated the transmission of natural images over an \gls{awgn} channel model with a finite channel input alphabet. Despite a good fit between the learned constellation diagram and the latent representation, the irregularity of the constellation diagram still poses significant challenges for deployment on commercial hardware. The author of \cite{Hu_2024_DTO} developed a digital task-oriented communication framework employing a hardware-limited scalar quantization approach, specifically tailored for computation-constrained situations, such as \gls{iot}. The results of this work provide valuable insights for future task-oriented \gls{jscc} designs.

\subsubsection{Edge Inference}
Edge inference has gained prominence as a solution to meet the stringent latency requirements of modern applications, which are not adequately supported by traditional cloud services \cite{Shi_2020_CEE, Li_2018_EIO}. The key architectural approach that underpins recent advancements is \textit{split inference}, where the inference network is partitioned between the device and the edge \cite{Huang_2020_DCR, Shi_2019_IDE, Li_2020_EAO, Shao_2020_CCT, Shao_2020_BAE, Jankowski_2020_JDE, Jankowski_2021_WIR, Shao_2022_LTO, Shao_2023_TOC}.

In this architecture, a mobile device initially processes data using a lightweight neural network to extract a compact feature vector. Subsequently, this vector is transmitted to an edge server for further processing, where deep \gls{jscc} is integral to the entire procedure \cite{Shao_2020_CCT, Shao_2020_BAE, Jankowski_2020_JDE, Jankowski_2021_WIR, Shao_2022_LTO, Shao_2023_TOC}. Notably, an end-to-end framework that efficiently compresses intermediate features to optimize the bandwidth and computational resources at the edge was introduced in \cite{Shao_2020_BAE}. In addition, the authors of \cite{Shao_2022_LTO} developed a method to flexibly adjust the length of the transmission signal to adapt to dynamic communication environments while maintaining targeted inference accuracy.

Recent studies have shifted from reconstruction-oriented communication, which focuses on accurately reconstructing a signal at the receiver, to a task-oriented approach that prioritizes inference accuracy as the primary performance metric \cite{Dubois_2021_LCf, Shao_2021_BGA, Shao_2022_LTO, Shao_2020_BAE, Shao_2023_TOC}. This paradigm shift underscores a move towards optimizing communication systems to support specific functional requirements rather than general data fidelity.

Note that implementing such split-design architectures often necessitates modifications on both the device and the edge, which pose challenges in terms of compatibility with existing communication infrastructures. This issue highlights a significant barrier to widespread adoption, indicating the need for more compatible solutions that can seamlessly integrate with current technologies.

\subsubsection{Variational Information Bottleneck}
The \gls{ib} theory, which extends from the foundational rate-distortion theory \cite{Cover_1991_EoI}, aims to find an optimal trade-off by maximizing the preservation of task-specific information in the latent representations, while minimizing the inclusion of task-agnostic information from the input data. Initially proposed by \cite{Tishby_1999_TIB}, the practical application of \gls{ib} theory in training deep neural networks remained theoretical until significantly later \cite{Tishby_2015_Dla}.

The application of \gls{ib} theory in deep learning was primarily hindered by computational challenges. The traditional optimization of the \gls{ib} objective function relied on the iterative Blahut-Arimoto algorithm \cite{Arimoto_1972_Aaf, Blahut_1972_Coc}, which is infeasible for deep learning applications due to its computational complexity and inefficiency in handling large-scale data \cite{Tishby_2015_Dla}. Addressing this limitation, \cite{Alemi_2017_DVI} introduced a variational approach to construct a tractable lower bound on the \gls{ib} objective, leading to the development of the \gls{vib} method. This approach enabled the practical application of the \gls{ib} principles in deep learning by approximating the intractable true posterior with a variational distribution.

Recent work has seen the integration of \gls{vib} with deep \gls{jscc}, which has been effectively used to formalize task-oriented communication strategies. In particular, the results \cite{Shao_2022_LTO, Shao_2023_TOC} have demonstrated that combining \gls{vib} with deep \gls{jscc} offers superior performance over reconstruction-oriented communication frameworks. These studies showcase the potential of \gls{vib} in improving the efficiency and robustness of communication systems, particularly in scenarios where preserving task-specific information and discarding task-agnostic information are crucial.

Integrating \gls{jscc} and \gls{ib} methods to protect user privacy is an advanced direction in current research. FedSem \cite{Wei_2023_FSL} had collaboratively trained semantic-channel encoders of multiple devices coordinated by a semantic-channel decoder using \gls{ib} theory based on base stations. Unlike traditional centralized learning approaches, FedSem reduces communication overhead and mitigates privacy concerns by enabling the sharing of semantic features rather than raw data. In addition, the author of \cite{Sun_2024_DIB} introduced a privacy-preserving \gls{jscc} scheme for image transmission, using a disentangled \gls{ib} objective to effectively separate private information from public data. This approach ensures the protection of privacy-sensitive information while maintaining high image quality. Although these works show impressive progress in the integration of \gls{jscc} with \gls{ib} theory, they often require specialized designs that are challenging to combine with existing systems and devices.

There is a need to design an advanced framework aligning two communication paradigms -- task-oriented communications and reconstruction-oriented communications -- and develop a \gls{jscc} modulation scheme for practical deployment.

\subsection{Contributions}
This paper introduces a novel communication framework compatible with reconstruction-oriented communication, especially for edge inference, termed \glsreset{atroc}\gls{atroc}. By extending \gls{ib} theory \cite{Tishby_1999_TIB} and incorporating \gls{jscc} modulation, this framework is designed to enhance AI-driven applications. It prioritizes task relevance in data transmission strategies, shifting focus from traditional signal reconstruction fidelity to operational efficiency and effectiveness in real-world applications. The key contributions of this research are summarized as follows:

\begin{itemize}
    \item \textbf{Development of an \gls{atroc} Framework:} Based on \gls{ib} theory, we develop a framework that aligns task-oriented communications with reconstruction-oriented communications. The framework focuses on maximizing mutual information between inference results and encoded features, minimizing mutual information between the encoded features and the input data, and preserving task-specific information through the information reshaper. This reshaper is expert at transforming received symbols into task-specific data, maintaining the same data structure as the input while ensuring the preservation of task-specific information.

    \item \textbf{Innovation of an Information Reshaper:} We introduce an information reshaper within our extended \gls{ib} theory, laying a foundational aspect of \gls{atroc}. This component is crucial for adapting the communication to the specific needs of the task without compromising the integrity of the transmitted data.

    \item \textbf{Variational Approximation for Tractable Information Estimation:} Due to the intractability of mutual information in the training and inference of deep neural networks, we employ a variational approximation approach, known as \gls{vib}. This approach allows us to establish a tractable upper bound for these terms, enabling training and inference of deep neural networks.

    \item \textbf{Adaptation of a \gls{jscc} Modulation Scheme:} We design a \gls{jscc} modulation scheme that aligns \gls{jscc} symbols with a predefined constellation scheme. This scheme ensures compatibility of our framework with classic modulation techniques, making it more adaptable to existing communication infrastructures.

    \item \textbf{Performance Enhancement in Edge-Based Autonomous Driving:} In our simulation, we validate that the \gls{atroc} framework outperforms reconstruction-oriented methods for edge-based autonomous driving \cite{Wu_2022_TgC}. 
    Specifically, our method reduced 99.19\% communication load, in terms of bits per service, compared to existing methods, without compromising the driving score of the autonomous driving agent.
\end{itemize}

\subsection{Organization and Notations}
The rest of this paper is organized as follows: \Cref{sec:aligned_TOC_framework} details the system model and discusses how the proposed framework advances reconstruction-oriented and non-aligned task-oriented communication approaches. \Cref{sec:AIB} introduces the \gls{ib} theory for \gls{atroc} and elaborates on the corresponding \gls{vib} derivation. In \cref{sec:modulation}, we propose a \gls{jscc} modulation technique that is compatible with classical modulation methods, such as \gls{qam}. \Cref{sec:edge_AI} extends the framework of \gls{vib} to enhance edge-based autonomous driving applications. The experimental results are presented in \cref{sec:result}, which evaluates the performance of our proposed \gls{atroc} framework and the \gls{jscc} modulation. Finally, \cref{sec:conclusion} concludes the paper.

\Cref{tab_notation} lists the main symbols used throughout this paper. 


\begin{table}
\centering
\footnotesize
\caption{SUMMARY OF MAIN SYMBOLS}
\label{tab_notation}
\begin{tabular}{ll} 
\toprule
\multicolumn{1}{c}{\textbf{Symbol}}                                                         & \multicolumn{1}{c}{\textbf{Explanation}}   \\ 
\midrule
$\bm{x}$                                                                                    & Input data                                 \\
$\hat{\bm{x}}$                                                                              & Reconstructed input data                   \\
$\bm{z}$                                                                                    & JSCC symbols                               \\
$\bar{\bm{z}}$                                                                              & Quantized JSCC symbols                     \\
$\bm{z}_{\text{in}}$                                                                        & Channel input     \\
$\bm{z}_{\text{out}}$                                                                       & Channel output                             \\
$\check{\bm{z}}$                                                                            & Equalized JSCC symbols                     \\
$\tilde{\bm{z}}$                                                                            & Scaled JSCC symbols                        \\
$\hat{\bm{z}}$                                                                              & Reconstructed JSCC symbols                 \\
$\bm{y}$                                                                                    & Task-specific data                         \\
$\bm{a}$                                                                                    & Target action            \\
$\hat{\bm{a}}$                                                                              & Inferred action                           \\
$\beta_1, \beta_2, \hat{\beta}_1, \hat{\beta}_2$                                            & Lagrange multiplier                        \\
$\phi, \theta, \psi, \delta$                                                                & Parameters of neural networks              \\
$h$                                                                                         & Channel coefficient                        \\
$\bm{n}$                                                                                    & Gaussian noise                             \\
$k$                                                                                         & Dimension of the JSCC symbols              \\
$l$                                                                                         & Dimension of the input data                \\
$\zeta$                                                                                     & Upper bound of rate                        \\
$\Omega$                                                                                    & Size of mini-batch                         \\
$u$                                                                                         & Number of constellation points             \\
$r$                                                                                         & Constellation parameter                    \\
$e_{\left( \cdot \right)}$                                                                    & Constellation point                        \\
$P_\text{target}$                                                                           & Power constraint of transmitter            \\
$P_{\bar{\bm{z}}}$                                                                            & Power of quantized symbols                 \\
$\beta_Q$                                                                                   & Hyperparameter of quantization loss        \\
$\Gamma, \lambda_\text{feat}, \lambda_\text{traj}, \lambda_\text{ctrl}, \lambda_\text{aux}$ & Hyperparameters of edge AI agent           \\
$J_1, J_2$                                                                                  & Sampling number                            \\
$i, j$                                                                                      & General index depended on context  \\
\bottomrule
\end{tabular}
\end{table}
\section{ATROC Framework for Edge Intelligence}
\label{sec:aligned_TOC_framework}
\begin{figure}[t]
    \begin{center}
    \includegraphics[width=1\linewidth]{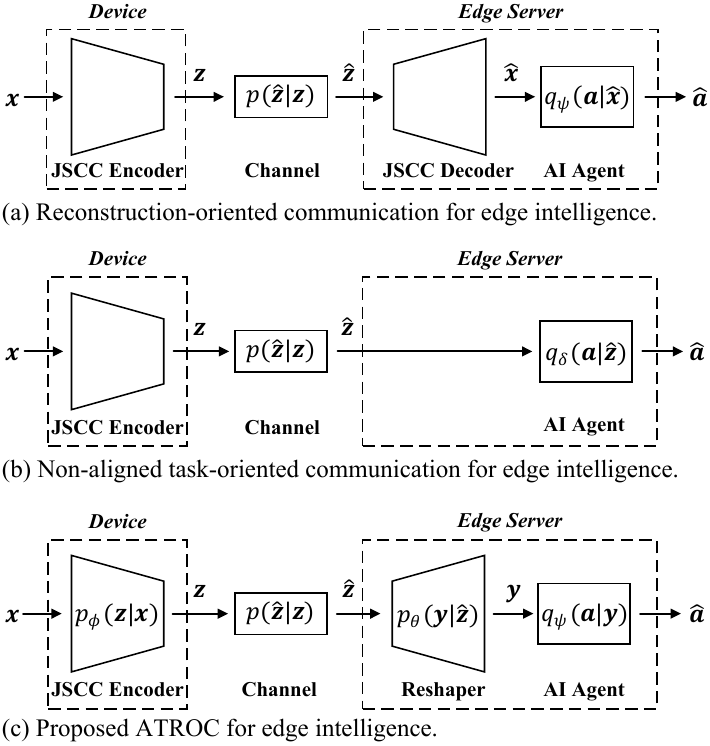}
    \end{center}
    \caption{Comparison of three JSCC-enabled communication frameworks for edge inference: Reconstruction-oriented, non-aligned task-oriented, and ATROC frameworks. All three frameworks can share a similar JSCC encoder structure on the device side. On the edge side, reconstruction-oriented communication aims to fully reconstruct the input data, including both task-specific and task-agnostic information. In contrast, non-aligned task-oriented communication focuses solely on preserving task-specific information and uses JSCC symbols directly for inference. ATROC merges the benefits of the previous two by transferring task-specific information and ensuring that data structures are compatible with existing AI agent networks, enhancing integration and efficiency.}
    \label{fig:IB_framework}
\end{figure}
Edge intelligence refers to an AI agent (system) that operates at edge servers rather than relying on centralized servers or cloud-based services. These systems process data locally on devices or at the edge of the network as shown in \cref{fig:IB_framework}. 

The reconstruction-oriented communication framework (see \cref{fig:IB_framework}a) aims to preserve all information from the input data $\bm{x}$ in the reconstructed data $\hat{\bm{x}}$. The idea is to minimize the distance $d(\bm{x}, \hat{\bm{x}})$, where $d(\cdot,\cdot)$ is a predefined data-centric metric. This task-agnostic strategy may result in transmitting redundant data for AI agents, leading to poor resource utilization efficiency.

To improve efficiency, the principle of \gls{ib} has been developed to transmit task-relevant information \cite{Shao_2022_LTO}, as shown in \cref{fig:IB_framework}b. 
However, a significant challenge arises with this approach: the dimensions of the received symbols often do not align with the input dimensions required by the edge AI agent. This mismatch necessitates a redesign of the AI agent to accommodate different input sizes, leading to poor compatibility.

To address this, we propose an \gls{atroc} framework, as depicted in \cref{fig:IB_framework}c, enabling the use of a unified inference network across both task-oriented and reconstruction-oriented communication paradigms.

In this framework, the \gls{jscc} encoder deployed on the mobile device is denoted by $p_\phi(\bm{z}|\bm{x})$, where $\phi$ represents the parameters. The encoder maps the input data $\bm{x} \in \mathbb{R}^{l}$ to \gls{jscc} symbols $\bm{z} \in \mathbb{C}^{k}$, where $\bm{z}=[z_1, \cdots, z_{k}]$. Here, $l$ and $k$ are the dimensions of the input data and the \gls{jscc} symbols, respectively. After quantization and power normalization, the \gls{jscc} symbols $\bm{z}$ are transmitted through a physical channel. In this paper, we model the communications between the mobile device and the edge server as Gaussian or Rayleigh fading channels:
\begin{align}
    \bm{z}_{\text{out}} = h \cdot \bm{z}_{\text{in}} + \bm{n},
    \label{eq_channel}
\end{align}
where $\bm{z}_{\text{in}}$ represents channel input and $\bm{z}_{\text{out}}$ represents channel output. $\bm{n} \sim \mathcal{CN}(0, \bm{\sigma}_{n}^{2} I)$ is a Gaussian noise with zero mean and standard deviation $\bm{\sigma}_{n}$. For the Gaussian channel, we set $h=1$, whereas for the Rayleigh fading channel, $h$ is modeled as a complex Gaussian variable, $h \sim \mathcal{CN}(0, 1)$, to represent the multipath fading effect.

After the process of equalization, scaling, and detection, the reconstructed symbols $\hat{\bm{z}}$ are transformed by the information reshaper $p_{\theta}(\bm{y}|\hat{\bm{z}})$ with parameters $\theta$ to provide task-specific data $\bm{y}$. These data are then utilized by the AI agent $q_\psi(\bm{a}|\bm{y})$ with parameters $\psi$, to generate the inferred action $\hat{\bm{a}}$, which approximates the ground truth action $\bm{a}$.

\section{Information Bottleneck for ATROC}
\label{sec:AIB}
\subsection{Problem Description}
Following the standard \gls{ib} framework \cite{Alemi_2017_DVI, Tishby_1999_TIB}, we assume the joint distribution of the system variables as follows:
\begin{equation}
    p(\bm{a}, \bm{x}, \bm{z}, \hat{\bm{z}}, \bm{y}) = p(\bm{a})p(\bm{x}|\bm{a})p_{\phi}(\bm{z}|\bm{x})p(\hat{\bm{z}}|\bm{z})p_{\theta}(\bm{y}|\hat{\bm{z}}).
\end{equation}
This sets up the Markov chain depicted as:
\begin{align}
    A \leftrightarrow X \leftrightarrow Z \leftrightarrow \hat{Z} \leftrightarrow Y.
\end{align}
We introduce a performance metric \textit{bits per service} to measure communication efficiency, which is defined as $k\cdot c$, where $c$ represents bits per symbol. Thus, there exists a crucial trade-off between bits per service and inference accuracy. This relationship underpins the formulation of an \gls{ib}, where we seek to optimize the balance between information throughput and decision accuracy.

The transformation from reconstructed symbols $\hat{\bm{z}}$ to task-specific data $\bm{y}$ is designed to preserve task-specific information, aligning task-oriented paradigms with traditional and reconstruction-oriented approaches. Based on the \gls{ib} theory \cite{Tishby_1999_TIB, Alemi_2017_DVI}, we formulate the following optimization problem:
\begin{subequations}
    \begin{align}
    \min \quad&-I(A;Y) \\
    \text{s.t.} \quad&I(X;\hat{Z})-\zeta \leq 0, \\
    &I(A;Y) - I(A;\hat{Z})= 0, \label{eq_optim_data_pro}
    \end{align}
\end{subequations}
where $\zeta$ represents the upper bound of data rate depending on the channel. The data processing inequality \cite{Cover_1991_EoI} implies that, ideally, if $Y$ and $\hat{Z}$ contain equivalent information about the action $A$, the equality $I(A;Y) - I(A;\hat{Z})=0$ holds.

\begin{figure*}
    \begin{subequations}
    \begin{align}
    \mathcal{L}_{\text{IB}}(\bm{a}, \bm{x}; \phi,\theta)&= \underbrace{-I(A;Y)}_{\text{Distortion}}
+\beta_1(\underbrace{I(X;\hat{Z})}_{\text{Rate}}-\zeta) +\beta_2\underbrace{(I(A;Y)-I(A;\hat{Z}))}_{\text{Alignment}} \label{eq_IB_origin}\\
&\equiv -I(A;Y) + \hat{\beta}_{1}I(X;\hat{Z})- \hat{\beta}_{2}I(A;\hat{Z}) \label{eq_IB_beta_hat}\\
&\equiv \mathbb{E}_{\bm{a}, \bm{x}}[\mathbb{E}_{\bm{y}|\bm{x};\phi,\theta}[-\log p(\bm{a}|\bm{y})]
+\hat{\beta}_{1} D_{K L}(p_{\phi}(\hat{\bm{z}} | \bm{x}) \| p(\hat{\bm{z}})) 
+\hat{\beta}_{2} \mathbb{E}_{\hat{\bm{z}} \mid \bm{x};\phi}[-\log p(\bm{a}|\hat{\bm{z}})]] \label{eq_IB_simplify}
    \end{align}
    \label{eq_IB}
    \end{subequations}
    {\noindent} \rule[0pt]{17.8cm}{0.05em}
\end{figure*}

We further formulate this problem as \cref{eq_IB_origin}, where $\beta_1 > 0$ and $\beta_2 > 0$ are the Lagrange multipliers. The detailed derivation can be found in \cref{subsec:VIB}.
The first term $-I(A;Y)$ and the second term $I(X;\hat{Z})$ formalize the classic information bottleneck, meanwhile, the third term $[I(A;Y)-I(A;\hat{Z})]$ aligns the task-relevant information between the task-specific data $\bm{y}$ and the reconstructed symbols $\hat{\bm{z}}$.

In the case $\beta_2 \neq 1$, we define $\hat{\beta}_{1} = \frac{\beta_1}{1-\beta_2}$ and $\hat{\beta}_{2} = \frac{\beta_2}{1-\beta_2}$. Then \cref{eq_IB_origin} can be expressed as \cref{eq_IB_beta_hat}. In the case $\beta_2 = 1$, \cref{eq_IB_origin} is simplified to the classic \gls{ib} formulation \cite{Tishby_1999_TIB, Alemi_2017_DVI, Shao_2022_LTO}:
\begin{align}
    \mathcal{L}_{\text{IB}}(\bm{a}, \bm{x}; \phi,\theta)=& \underbrace{-I(A;\hat{Z})}_{\text{Distortion}}
+\beta_1\underbrace{I(X;\hat{Z})}_{\text{Rate}}. 
\end{align}
This extended \gls{ib} theory preserves more task-specific information, and the bits per service is the same as the previous \gls{ib} approaches. Meanwhile, it maintains the dimension and structure required for edge inference.

\subsection{Variational Approach}
\label{subsec:VIB}
With the objective function \cref{eq_IB_beta_hat}, we first illustrate how to compute each term for training \(\phi\) and \(\theta\). We start with the first term, $-I(A;Y)$, expressed as:
\begin{align}
    -I(A;Y)=& -\int p(\bm{a},\bm{y})\log{\frac{p(\bm{a}|\bm{y})}{p(\bm{a})}} \dif \bm{a} \dif \bm{y} \notag\\
    =& -\int p(\bm{a},\bm{y})\log{p(\bm{a}|\bm{y})} \dif \bm{a} \dif \bm{y} - H(A),
\end{align}
where $p(\bm{a}|\bm{y})$ is the posterior probability, which can be derived through the Markov Chain \cite{Alemi_2017_DVI, Shao_2022_LTO} as:
\begin{align}
    p(\bm{a}|\bm{y}) &= \int p(\bm{a},\bm{x},\bm{z},\hat{\bm{z}}|\bm{y})\dif{\bm{x}}\dif{\bm{z}}\dif{\hat{\bm{z}}} \notag\\
    =& \int\frac{p(\bm{a})p(\bm{x}|\bm{a})p_{\phi}(\bm{z}|\bm{x})p(\hat{\bm{z}}|\bm{z})p_{\theta}(\bm{y}|\hat{\bm{z}})}{p(\bm{y})}\dif{\bm{x}}\dif{\bm{z}}\dif{\hat{\bm{z}}}.
\end{align}
Given the complexity of this integration, we employ a neural network $q_{\psi}(\bm{a}|\bm{y})$ as a variational approximation to $p(\bm{a}|\bm{y})$. 

Denoting the \gls{kl} divergence as $D_{\text{KL}}$. According to the definition of \gls{kl} divergence \cite{Cover_1991_EoI}, we can derive the following expression:
\begin{align}
D_{\text{KL}}(p(\bm{a}|\bm{y}) \parallel &q_{\psi}(\bm{a}|\bm{y})) \notag\\
=&\int p(\bm{a},\bm{y})\log p(\bm{a}|\bm{y}) \dif\bm{a}\dif\bm{y} \notag\\
&- \int p(\bm{a},\bm{y}) \log q_{\psi}(\bm{a}|\bm{y}) \dif\bm{a}\dif\bm{y}.
\end{align}
Based on the fact that 
\begin{align}
    D_{\text{KL}}(p(\bm{a}|\bm{y}) \parallel q_{\psi}(\bm{a}|\bm{y})) \geq 0,
    \label{eq_KL_conditional}
\end{align} 
we have
\begin{align}
\int p(\bm{a},\bm{y})\log p(\bm{a}|\bm{y})& \dif\bm{a}\dif\bm{y} \notag\\
\geq &\int p(\bm{a},\bm{y})\log q_{\psi}(\bm{a}|\bm{y}) \dif\bm{a}\dif\bm{y},
\label{eq_KL_conditional_2}
\end{align}
which derives
\begin{align}
\mathbb{E}_{\bm{a},\bm{x}}\bigl[\mathbb{E}_{\bm{y}|\bm{x};\phi,\theta}&[-\log p(\bm{a}|\bm{y})]\bigr] \notag\\
&\leq \mathbb{E}_{\bm{a},\bm{x}}\left[\mathbb{E}_{\bm{y}|\bm{x};\phi,\theta}[-\log q_{\psi}(\bm{a}|\bm{y})]\right].
\label{eq_firstTerm_mean}
\end{align}
The detailed derivation of \cref{eq_firstTerm_mean} can be found in Appendix \ref{apd_fristTerm_mean}.

The second term $I(X;\hat{Z})$ \cite{Shao_2022_LTO} is formulated as:
\begin{align}
I(X;\hat{Z})=\mathbb{E}_{\bm{a},\bm{x}}\left[D_{\text{KL}}(p_{\phi}(\hat{\bm{z}} | \bm{x}) \| p(\hat{\bm{z}})) \right],
\end{align}
where the marginal probability is 
\begin{align}
    p(\hat{\bm{z}})=\int p(\bm{a})p(\bm{x}|\bm{a})p_{\phi}(\bm{z}|\bm{x})p(\hat{\bm{z}}|\bm{z}) \dif\bm{a}\dif\bm{x}\dif\bm{z}.
\end{align}
We adopt a Gaussian approximation $q(\hat{\bm{z}}) \sim \mathcal{N}(\bm{0}, I)$ as an estimation for $p(\hat{\bm{z}})$ \cite{Kingma_2013_Aev}. It is reasonable as the \gls{jscc} encoder generates a Gaussian distribution $p_{\phi}(\hat{\bm{z}}|\bm{x}) \sim \mathcal{N}(\bm{\mu}_{\phi}(\bm{x}), \bm{\sigma}_{\phi}^2(\bm{x})I)$, where $\bm{\mu}_{\phi}(\cdot)$ and $\bm{\sigma}_{\phi}(\cdot)$ are functions that map the input data $\bm{x}$ to the mean and standard deviation of the Gaussian distribution.

Since $D_{\text{KL}}(p(\hat{\bm{z}})\parallel q(\hat{\bm{z}})) \geq 0$, the following upper bound can be derived:
\begin{align}
    I(X;\hat{Z}) \leq \mathbb{E}_{\bm{a},\bm{x}}\left[D_{\text{KL}}(p_{\phi}(\hat{\bm{z}} | \bm{x}) \| q(\hat{\bm{z}})) \right],
\end{align}
where the \gls{kl} divergence can be calculated analytically by the method in \cite{Duchi_2007_DfL}.

Similar to \cref{eq_firstTerm_mean}, by using $q_{\delta}(\bm{a}|\hat{\bm{z}})$ as a variational approximation of $p(\bm{a}|\hat{\bm{z}})$, we have
\begin{align}
    \mathbb{E}_{\bm{a},\bm{x}}\bigl[\mathbb{E}_{\hat{\bm{z}}|\bm{x};\phi,\theta}[-&\log p(\bm{a}|\hat{\bm{z}})]\bigr] \notag\\
    &\leq \mathbb{E}_{\bm{a},\bm{x}}\left[\mathbb{E}_{\hat{\bm{z}}|\bm{x};\phi,\theta}[-\log q_{\delta}(\bm{a}|\hat{\bm{z}})]\right].
\end{align} 
The above extended \gls{vib} formulation determines the upper bound of the \gls{ib} objective function (\cref{eq_IB_simplify}), which can be expressed as:
\begin{align}
\mathcal{L}_{\text{VIB}}(\bm{a}, \bm{x};\phi, \theta)= \mathbb{E}_{\bm{a},\bm{x}}&\Bigl\{
\mathbb{E}_{\bm{y}|\bm{x};\phi,\theta}[-\log q_{\psi}(\bm{a}|\bm{y})] \notag\\
&+\hat{\beta}_1 D_{\text{KL}}(p_{\phi}(\hat{\bm{z}} | \bm{x}) \| q(\hat{\bm{z}})) \notag\\
&+\hat{\beta}_2 \mathbb{E}_{\hat{\bm{z}}|\bm{x};\phi,\theta}[-\log q_{\delta}(\bm{a}|\hat{\bm{z}})]
\Bigr\}.
\label{eq_VIB_theory}
\end{align}
Through Monte Carlo sampling, we train \(\phi\) and \(\theta\) by minimizing this objective function using stochastic gradient descent.
Specifically, given a mini-batch of data $\{(\bm{a}_i,\bm{x}_i)\}^\Omega_{i=1}$ with batch size $\Omega$, if the reconstructed \gls{jscc} symbols $\hat{\bm{z}}$ are sampled $J_1$ times and the task-specific data $\bm{y}$ are sampled $J_2$ times for each data pair, the following estimation can be obtained:
\begin{align}
    \mathcal{L}_{\text{VIB}}(\bm{a}, \bm{x};\phi, \theta)\cong \frac{1}{\Omega}\sum_{i=1}^{\Omega} 
&\left\{
\frac{1}{J_2}\sum_{j=1}^{J_2}[-\log q_{\psi}(\bm{a}_{i}|\bm{y}_{j})] \right. \notag\\ 
&\left. +\hat{\beta}_1 D_{\text{KL}}(p_{\phi}(\hat{\bm{z}} | \bm{x}_{i}) \| q(\hat{\bm{z}})) \right. \notag\\
&\left. +\frac{\hat{\beta}_2}{J_1}\sum_{j=1}^{J_1}[-\log q_{\delta}(\bm{a}_{i}|\hat{\bm{z}}_{j})]
\right\}.
\label{eq_VIB_sampling}
\end{align}

\section{JSCC Modulation}
\label{sec:modulation}
\begin{figure*}[t]
    \begin{center}
    \includegraphics[width=1\linewidth]{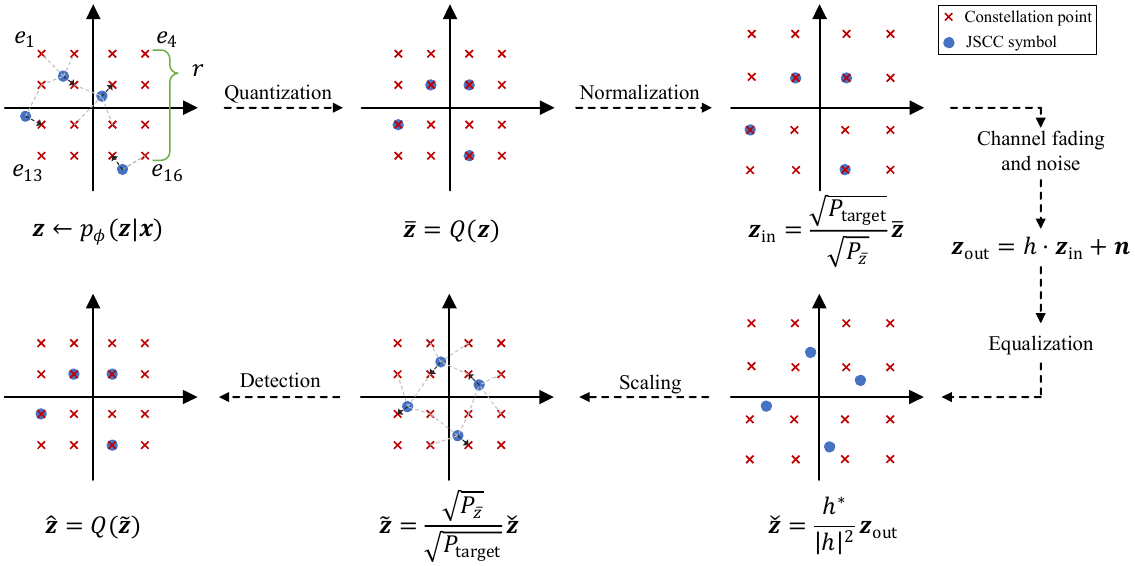}
    \end{center}
       \caption{An example of the JSCC modulation and signal transmission procedure for $\bm{z} \in \mathbb{C}^4$ using 16-QAM.}
    \label{fig:Quantization}
\end{figure*}

In existing communication standards, symbols are transmitted with specific constellation orders and designs.  
In this section, we develop a \gls{jscc} modulation scheme that can map arbitrary complex-valued \gls{jscc} symbols to a predefined constellation diagram with finite points, as shown in \cref{fig:Quantization}. In addition, we introduce a learning method to adjust the optimal constellation parameter according to the quantization loss. For clarity, we use \gls{qam} as an example. Note that our method can be easily extended to other modulation schemes.

\subsection{Quantization and Normalization}
To enable the quantization of arbitrary complex-valued \gls{jscc} symbols into a predefined constellation diagram, the following rule is applied to each symbol:
\begin{align}
    \bar{z}_i=Q(z_i)=\arg\mathop{\min}_{e_j}\|z_i-e_j \|_2^2,
    \label{eq_quantization}
\end{align}
where $z_i \in \mathbb{C}$ represents the original symbol, $\bar{z_i} \in \mathbb{C}$ represents the quantized symbol, $i\in \{1, \cdots,k \}$, $Q(\cdot)$ denotes the quantization function, and $\|\cdot \|_2$ denote the $\ell^2$-norm. $e_j \in \{e_1, \cdots,e_u\}$ represents the predefined constellation points, where $e_j\in \mathbb{C}$, and $u$ denote the number of constellation points. This quantization operation can be extended to a vector as follows,
\begin{align}
    \bar{\bm{z}}=Q(\bm{z})=[Q(z_1), \cdots, Q(z_k)].
    \label{eq_quantization_vector}
\end{align}

Since the transmitted symbols should satisfy the average power constraint:
\begin{align}
    \frac{1}{k}\sum_{i=1}^{k} |\bar{z}_i|^2 \leq P_{\text{target}},
\end{align}
the channel input (normalized symbols) are given by:
\begin{align}
\bm{z}_\text{in} = \frac{\sqrt{P_{\text{target}}}}{\sqrt{P_{\bar{\bm{z}}}}}\cdot\bar{\bm{z}},
\end{align}
where $P_{\bar{\bm{z}}} = \frac{1}{k} \sum_{i=1}^{k} |\bar{z}_i|^2$ denotes the power of quantized symbols $\bar{\bm{z}}$.

The channel input $\bm{z}_\text{in}$ is transmitted through the channel $\bm{z}_{\text{out}} = h\cdot\bm{z}_{\text{in}} + \bm{n}$.
Assume that the receiver has the full \gls{csi} knowledge and knows $P_{\bar{\bm{z}}}$, in the case of the static channel, it can perform channel equalization:
\begin{align}
    \check{\bm{z}} = \frac{h^*}{|h|^2}\bm{z}_{\text{out}},
\end{align}
where $h^*$ denotes the conjugate of channel coefficient $h$ and $\check{\bm{z}}$ denotes the equalized symbols.  After equalization, the equalized symbols should be scaled as
\begin{align}
    \tilde{\bm{z}} = \frac{\sqrt{P_{\bar{\bm{z}}}}}{\sqrt{P_{\text{target}}}}\cdot\check{\bm{z}},
\end{align}
where $\tilde{\bm{z}}$ denotes the scaled symbols. Then the reconstructed symbols $\hat{\bm{z}}=Q(\tilde{\bm{z}})$ can be obtained by \cref{eq_quantization_vector}.

\subsection{Learnable Constellation Diagram and Fine-Tuning}
\label{subsec:constellation}
Traditional modulation techniques, such as \gls{qam}, employ a lookup table that maps bits to constellation points. In contrast, the complex-valued channel symbols produced by the \gls{jscc} encoder are continuous, necessitating a different approach for their mapping.

\Cref{eq_quantization} demonstrates that the coordinates of each constellation point $e_j$ directly affect the quantization outcome. We propose a learnable constellation diagram that adapts to the observed space of \gls{jscc} symbols, minimizes quantization loss, and improves performance with the \gls{jscc} encoder and the information reshaper. Taking $u$-\gls{qam} as an example, where $u$ denotes the number of constellation points, the coordinates of each constellation point can be derived by the parameter $r$. This parameter specifies the distance between two constellation points located at the corners of one side, as illustrated in \cref{fig:Quantization}. Then, the real part and imaginary part of the constellation point \(e_j\) are given as follows:
\begin{align}
\Re(e_{j})=&-\frac{r}{2}+\frac{(j\bmod\sqrt{u})\cdot r}{\sqrt{u}-1}, \\
\Im(e_{j})=&\frac{r}{2}-\frac{\lfloor{j/\sqrt{u}}\rfloor\cdot r}{\sqrt{u}-1},
\end{align}
where ``\(\,\bmod\,\)'' denotes the modulo operation and $\lfloor\cdot\rfloor$ denotes the rounding down function.

The quantization loss is defined as
\begin{align}
    \mathcal{L}_{Q}(\bm{z};r)=\frac{1}{k}\sum_{i=1}^{k}\mathop{\min}_{e_j}\|z_i-e_j \|_2.
\end{align}

The training process for the learnable constellation diagram begins with the initialization of the constellation parameter \(r\) to a predefined value \(r_{\text{init}}\), along with loading a pre-trained \gls{jscc} encoder. Using an image dataset \(\mathcal{X}\) with corresponding ground truth actions \(\mathcal{A}\), mini-batches are sampled iteratively during training. For each mini-batch, images are encoded into \gls{jscc} symbols, and the average batch loss is computed based on the quantization error. The constellation parameter \(r\) is then updated by backpropagation until convergence. The output of this process is the optimal constellation parameter \(r^*\). The detailed constellation parameter training process is provided in \cref{alg_quantization}. Once the optimal $r^*$ is obtained, the \gls{jscc} encoder and the information reshaper are jointly fine-tuned using the extended loss function:
\begin{align}
    \mathcal{L}_{\text{VIB-}Q}(\bm{a}, \bm{x};\phi,\theta)=\mathcal{L}_{\text{VIB}}(\bm{a}, \bm{x};\phi,\theta) + \beta_{Q}\mathcal{L}_{Q}(\bm{z};r^*),
    \label{eq_loss_VIBQ}
\end{align}
where $\beta_{Q}$ is a hyperparameter that balances the quantization loss with the original \gls{vib} loss.

This method enhances the practical applicability of \gls{jscc} modulation by integrating it with established digital communication systems while preserving the benefits of customized encoding and decoding strategies.

\begin{algorithm}[t]
\caption{Training Learnable Constellation Diagram}
\label{alg_quantization}
\begin{algorithmic}[1]
\Statex \textbf{Initialization}: Initialize the constellation parameter
\Statex $r\rightarrow r_{\text{init}}$, and load pre-trained \gls{jscc} encoder $p_{\phi}(\bm{z}|\bm{x})$.
\State \textbf{Input}: Image dataset $\mathcal{X}$ with corresponding ground truth action $\mathcal{A}$.

\While{not converged}
    \State Sample mini-batch $\{(\bm{a}_i, \bm{x}_i) \}_{i=1}^{\Omega}$ from $\mathcal{X}$ and $\mathcal{A}$.
    \State Encode image $\{\bm{x}_i\}_{i=1}^{\Omega}$ to symbols $\{\bm{z}_i\}_{i=1}^{\Omega}$.
    \State Compute the average batch loss
    \Statex \quad \; $\frac{1}{\Omega}\sum_{i=1}^{\Omega}\mathcal{L}_{Q}(\bm{z}_i;r)$.
    \State Update parameter $r$ through backpropagation.
\EndWhile
\State \textbf{Output}: Optimal constellation parameter $r^*$.

\end{algorithmic}
\end{algorithm}

\section{Extended VIB for Edge-based Autonomous Driving}
\label{sec:edge_AI}
\gls{tgcp}\footnote{To avoid confusion with the Transmission Control Protocol (TCP), we denote Trajectory-guided Control Prediction as TGCP in this paper.} is the state-of-the-art \gls{e2e} self-driving framework that combines trajectory planning and multi-stage control prediction into a unified neural network \cite{Wu_2022_TgC}. 
This framework, notable for using only a monocular camera, ranks third on the CARLA leaderboard\footnote{\url{https://leaderboard.carla.org/leaderboard/}}. 
We extend \gls{vib} to \gls{tgcp} to examine its applicability in an edge-based autonomous driving system.

\subsection{Background of TGCP}
\gls{tgcp} at the edge server processes task-specific data $\bm{y}$ and additional state information $\bm{m}$ to make driving decisions. The state information includes variables such as speed, destination coordinates, and current driving guidance (e.g., ``turn left'' or ``follow the lane''). For this study, we assume that $\bm{m}$ can be transmitted losslessly to the edge server. 

The autonomous driving agent is modeled as $q_{\psi}(\bm{a}|\bm{y})$, which generates the inferred action $\hat{\bm{a}}$ from task-specific data $\bm{y}$. In particular, the individual components of the inferred action $\hat{\bm{a}} = (\hat{v}, \hat{s}, \hat{\bm{w}}, \hat{\bm{f}}^{\text{traj}}, \hat{\bm{b}}, \hat{\bm{f}}^{\text{ctrl}})$ are defined as follows:
\begin{itemize}
    \item $\hat{v}$: estimated target speed.
    \item $\hat{s}$: value of the extracted features estimated by the expert \cite{Zhang_2021_EtE}.
    \item $\hat{\bm{w}}$: predicted waypoints from the trajectory branch.
    \item $\hat{\bm{f}}^{\text{traj}}$: estimated extracted features for trajectory planning.
    \item $\hat{\bm{b}} = [\hat{\bm{b}}_0, \hat{\bm{b}}_1, \dots, \hat{\bm{b}}_\Gamma]$: estimated control actions from the beta distribution in the control prediction branch, where $\Gamma$ denotes the prediction length.
    \item $\hat{\bm{f}}^{\text{ctrl}} = [\hat{\bm{f}}^{\text{ctrl}}_0, \hat{\bm{f}}^{\text{ctrl}}_1, \dots, \hat{\bm{f}}^{\text{ctrl}}_\Gamma]$: predicted informative features of the control prediction branch.
\end{itemize}

\subsection{Control and Trajectory Prediction Loss Functions}
The designed controller, based on \cite{Wu_2022_TgC}, computes control commands such as throttle, steer, and brake using the output of the trajectory and control prediction branches. The corresponding loss functions are defined as:
\begin{align}
    \mathcal{L}_{\text{traj}} =& \|\bm{w} - \hat{\bm{w}}\|_1 + \lambda_{\text{feat}}\|\bm{f}^{\text{traj}} - \hat{\bm{f}}^{\text{traj}}\|_2, \\
    \mathcal{L}_{\text{ctrl}}
    =& D_{\text{KL}}(\mathcal{B}e(\bm{b}_{0})\|\mathcal{B}e(\hat{\bm{b}}_{0}))  \notag\\
    &+\frac{1}{\Gamma}\sum_{i=1}^{\Gamma}D_{\text{KL}}(\mathcal{B}e(\bm{b}_{i})\|\mathcal{B}e(\hat{\bm{b}}_{i})) \notag\\
    & + \lambda_{\text{feat}}\|\bm{f}^{\text{ctrl}}_{0} - \hat{\bm{f}}^{\text{ctrl}}_{0}\|_{2}
    + \frac{1}{\Gamma}\sum_{i=1}^{\Gamma}\|\bm{f}^{\text{ctrl}}_{i} - \hat{\bm{f}}^{\text{ctrl}}_{i}\|_{2},
\end{align}
where $\lambda_\text{feat}$ is a hyperparameter, $\bm{w}$, $\bm{f}^{\text{traj}}$, $\bm{b}_{i}$, and $\bm{f}_{i}^{\text{ctrl}}$ are from the ground truth action $\bm{a}$, $\|\cdot\|_{1}$ denotes the $\ell_{1}$-norm, and $\mathcal{B}e(\cdot)$ denotes the beta distribution.

Furthermore, the auxiliary loss function is defined as:
\begin{equation}
    \mathcal{L}_{\text{aux}} = \|v-\hat{v}\|_{1} + \|s-\hat{s}\|_{2},
\end{equation}
where speed $v$ and value of features $s$ are from the ground truth action $\bm{a}$. Combining these terms, the overall loss function $\mathcal{L}_{\text{TCGP}}$ becomes:
\begin{equation}
    \mathcal{L}_{\text{TCGP}} = \lambda_{\text{traj}}\mathcal{L}_{\text{traj}} + \lambda_{\text{ctrl}}\mathcal{L}_{\text{ctrl}} + \lambda_{\text{aux}}\mathcal{L}_{\text{aux}},
\end{equation}
where $\lambda_{\text{traj}}$, $\lambda_{\text{ctrl}}$, and $\lambda_{\text{aux}}$ are hyperparameters.

\subsection{Task-Oriented End-to-End Training}
\label{subsec_task-oriented_training}
Typically, we assume that the posterior $q_{\psi}(\bm{a}|\bm{y})$ follows a Gaussian distribution $\mathcal{N}(\bm{\mu}_{\psi}(\bm{y}), \bm{\Sigma}_{\psi}(\bm{y}))$, where $\bm{\mu}_{\psi}(\bm{y})\in\mathbb{R}^{d}$ and $\bm{\Sigma}_{\psi}(\bm{y})=\sigma_{c}^{2}I_{d}$ ($\sigma_{c}$ is a constant). According to the probability density function of the Gaussian distribution, we can derive the following expression,
\begin{align}
    -\log{q_{\psi}(\bm{a}|\bm{y})}\sim \frac{1}{2\sigma^{2}_{c}}\|\bm{a}-\bm{\mu}_{\psi}(\bm{y}) \|^2_2,
    \label{eq_log2normal}
\end{align}
where $\bm{\mu}_{\psi}(\bm{y})=\hat{\bm{a}}$.
Details of the derivation are deferred to the Appendix \ref{apd:derivation_log}. \cref{eq_log2normal} shows that $-\log q_{\psi}(\bm{a}|\bm{y})$ can serve as a distance metric, like the $\ell^2$-norm. 
Since $\mathcal{L}_{\text{TCGP}}$ is a combination of distance metric of action $\bm{a}$ ($\ell^1$-norm, $\ell^2$-norm, and \gls{kl} divergence), we heuristically propose substituting the first term in \cref{eq_VIB_theory} with $\mathcal{L}_{\text{TCGP}}$ to adapt the objective function as:
\begin{align}
\mathcal{L}_{\text{VIB}}'(\bm{a}, \bm{x};\phi, \theta)=\mathbb{E}_{\bm{a},\bm{x}}&\Bigl\{
\mathcal{L}_{\text{TCGP}} \notag\\
&+\hat{\beta}_1 D_{\text{KL}}(p_{\phi}(\hat{\bm{z}} | \bm{x}) \| q(\hat{\bm{z}})) \notag\\
&+\hat{\beta}_2 \mathbb{E}_{\hat{\bm{z}}|\bm{x};\phi,\theta}[-\log q_{\delta}(\bm{a}|\hat{\bm{z}})]
\Bigr\}.
\label{eq_VIB_2}
\end{align}
In addition, the \cref{eq_loss_VIBQ} can be modified as:
\begin{align}
    \mathcal{L}_{\text{VIB-}Q}'(\bm{a}, \bm{x};\phi,\theta)=\mathcal{L}_{\text{VIB}}'(\bm{a}, \bm{x};\phi,\theta) + \beta_{Q}\mathcal{L}_{Q}(\bm{z};r^*).
    \label{eq_loss_VIBQ_2}
\end{align}

Training of \gls{jscc} encoder and information reshaper consists of two stages: pre-training and fine-tuning. In pre-training, the neural network parameters (\(\phi\) and \(\theta\)) are initialized, and images from the dataset are encoded into \gls{jscc} symbols, transmitted through a channel without modulation, and reshaped into task-specific data. The \gls{tgcp} model, with frozen parameters, generates inferred actions \(\hat{\bm{a}}\), and the loss \(\mathcal{L}_{\text{VIB}}'\) is computed to update the network parameters. Fine-tuning follows a similar process, but the symbols are transmitted with JSCC modulation, and the loss \(\mathcal{L}_{\text{VIB-}Q}'\) is used for parameter updates. Finally, the optimized parameters \(\phi\) and \(\theta\) are output. The training process of the proposed aligned task- and reconstruction-oriented \gls{jscc} encoder and information reshaper is shown in \cref{alg_training}. Here, \(\text{CH}(\cdot)\) denotes the function of a \gls{jscc} modulation and communication channel, while \(\text{TGCP}(\cdot)\) denotes the function of \gls{tgcp}. Specifically, during the fine-tuning process, both the JSCC encoder and the information reshaper are actively adjusted, which means that neither component is frozen. This fine-tuning process reduces the quantization loss of the encoder's output and preserves task-critical information, showing the potential for real-world applications.

\begin{algorithm}[t]
\caption{Training JSCC Encoder and Information Reshaper.}
\label{alg_training}
\begin{algorithmic}[1]

\Statex \textbf{Initialization}: Initialize the neural network parameters \(\phi\) and \(\theta\).

\State \textbf{Input}: Image dataset $\mathcal{X}$ with corresponding ground-truth agent output $\mathcal{A}$. Well-trained TGCP model with frozen parameters. Learning rate \(\eta\).

\While{not converged}
    \State Sample mini-batch $\{(\bm{a}_i, \bm{x}_i)\}_{i=1}^{\Omega}$ from $\mathcal{A}$ and $\mathcal{X}$.
    
    \For{sample $i=1,\dots,\Omega$}
        \State Encode image \(\bm{x}_{i}\) to JSCC symbols \(\bm{z}_{i}\).
        \State Transmit JSCC symbols through a channel  
        \Statex \qquad \ \ \ without JSCC modulation: \(\hat{\bm{z}}_i\leftarrow\text{CH}(\bm{z}_i)\).
        \State Reshape the reconstructed JSCC symbols \(\hat{\bm{z}}_i\)
        \Statex \qquad \ \ \ to task-specific data \(\bm{y}_i\).
        \State Generate inferred action: \(\hat{\bm{a}}_i\leftarrow\text{TGCP}(\bm{y}_i)\).
        \State Compute loss \(\mathcal{L}_{\text{VIB}}'\) based on \cref{eq_VIB_2}.
    \EndFor
    \State Update parameters (pre-training):
    \Statex \quad \, \(\phi \overset{+}\leftarrow -\eta\cdot\nabla_{\phi}\mathcal{L}_\text{VIB}' \), \(\theta \overset{+}\leftarrow -\eta\cdot\nabla_{\theta}\mathcal{L}_\text{VIB}' \).
\EndWhile
\State Find optimal constellation parameter \(r^{*}\) according to \cref{alg_quantization}.
\While{not converged}
    \State Sample mini-batch $\{(\bm{a}_i, \bm{x}_i)\}_{i=1}^{\Omega}$ from $\mathcal{A}$ and $\mathcal{X}$.
    
    \For{sample $i=1,\dots,\Omega$}
        \State Encode image \(\bm{x}_{i}\) to JSCC symbols \(\bm{z}_{i}\).
        \State Transmit JSCC symbols through a channel  
        \Statex \qquad \ \ \ with JSCC modulation: \(\hat{\bm{z}}_i\leftarrow\text{CH}(\bm{z}_i)\).
        \State Reshape the reconstructed JSCC symbols \(\hat{\bm{z}}_i\)
        \Statex \qquad \ \ \ to task-specific data \(\bm{y}_i\).
        \State Generate inferred action: \(\hat{\bm{a}}_i\leftarrow\text{TGCP}(\bm{y}_i)\).
        \State Compute loss \(\mathcal{L}_{\text{VIB-}Q}'\) based on \cref{eq_loss_VIBQ_2}.
    \EndFor
    \State Update parameters (fine-tuning):
    \Statex \quad \, \(\phi \overset{+}\leftarrow -\eta\cdot\nabla_{\phi}\mathcal{L}_{\text{VIB-}Q} '\), \(\theta \overset{+}\leftarrow -\eta\cdot\nabla_{\theta}\mathcal{L}_{\text{VIB-}Q} '\). 

\EndWhile

\State \textbf{Output}: Neural network parameters: \(\phi\) and \(\theta\).

\end{algorithmic}
\end{algorithm}

\section{Performance Evaluation}
\label{sec:result}
\subsection{Experiment Setup}
\subsubsection{Dataset}
We utilize the Car Learning to Act (CARLA) simulator, an open-source platform designed for autonomous driving research \cite{Dosovitskiy_2017_CAO}, which provides a variety of urban environments that simulate real-world traffic scenarios. The image dataset from \cite{Wu_2022_TgC}, comprising images from various urban maps, serves as the input data \(\bm{x}\) for our training. In our experiments, the training dataset consists of 189,524 images from four maps: Town01, Town03, Town04, and Town06. The test dataset includes 27,201 images from another four maps: Town02, Town05, Town07, and Town10.

\subsubsection{Evaluation Metrics}
Our evaluation focuses on comparing the driving performance of our \gls{atroc} framework against various baselines within the CARLA simulator. We use the commonly adopted driving score\footnote{\url{https://leaderboard.carla.org/}} to assess the vehicle's ability to navigate according to predetermined waypoints, destinations, and comply with traffic regulations. Each test is conducted three times in Town05 under four different weather scenarios: clear noon, cloudy sunset, soft rain at dawn, and hard rain at night.

\subsubsection{Basic Settings}
In our proposed framework, we configure the \gls{jscc} symbols dimension $k$ to $1024$, enabling us to achieve significantly low bits per service of $6144$ for 64-QAM. For training of the \gls{jscc} encoder and the information reshaper, we set the Lagrange multipliers $\hat{\beta}_1 = 1$, $\hat{\beta}_2 = 8192$, and the quantization loss hyperparameter $\beta_{Q}=10$, which is a good balance between fidelity and compression. Furthermore, we set the learning rate \(\eta\) to \(4\times10^{-5}\), and impose a power constraint with $P_{\text{target}} = 1$. 

The \gls{tgcp} model is trained following the instructions of \cite{Wu_2022_TgC}. In addition, the pre-trained parameters $\psi$ of \gls{tgcp} are kept fixed throughout all training phases. This pre-trained \gls{tgcp} model serves as the AI agent in the following experiments.

\begin{figure*}[t]
    \begin{center}
    \includegraphics[width=1.0\linewidth]{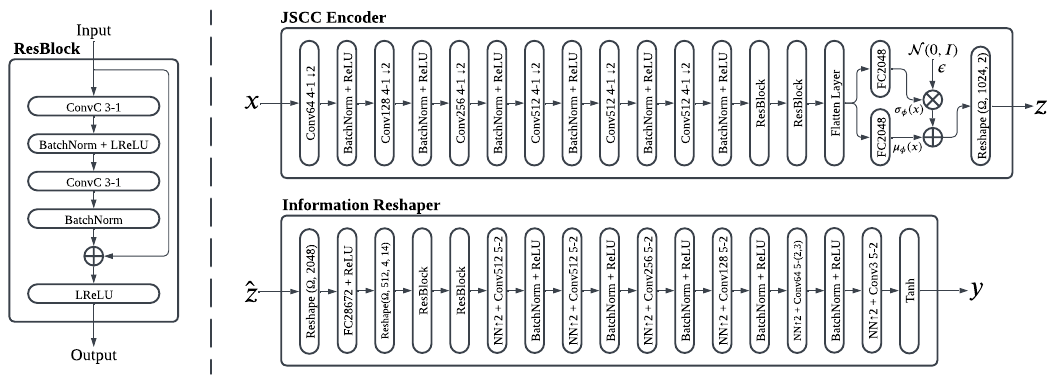}
    \end{center}
       \caption{Architecture of the proposed JSCC encoder and information reshaper. For example, \textit{ConvC 3-1} represents a convolutional layer with \(C\) channels, a \(3\times3\) kernel size, and padding of 1 on both sides. \(\downarrow\)2 denotes the strided down convolutions, while NN\(\uparrow\)2 denotes the nearest neighbor upsampling. \textit{FC2048} refers to a fully connected layer with an output size of 2048. \textit{BatchNorm} denotes batch normalization, \textit{LReLU} represents the leaky ReLU activation with \(\alpha=0.2\), and \(\Omega\) represents the batch size. The dimensions (number of channels) of the inputs and outputs for the \textit{ResBlock} remain unchanged.}
    \label{fig_codec_structure}
\end{figure*}

For simplicity, a deterministic information reshaper is used, allowing us to approximate $q_{\delta}(\bm{a}|\hat{\bm{z}})$ by $q_{\psi}(\bm{a}|\bm{y})$. The architecture and detailed parameters of the proposed JSCC encoder and the information reshaper are shown in \cref{fig_codec_structure}.

\subsubsection{Baseline Methods}
Three traditional image coding methods are included as baseline methods for comparison: (1) JPEG \cite{Wallace_1992_TJs}; (2) JPEG2000 \cite{Taubman_2002_JIc}; (3) and BPG \cite{BPG}. Each traditional image coding method is followed by (2048, 6144) \gls{ldpc} codes combined with a 64-QAM digital modulation scheme. The average bits per service for these methods range from 36,844 to 1,041,758. 

In addition, baseline methods also include two state-of-the-art reconstruction-oriented \gls{jscc} designs, with the legends ``ROC-AE'' \cite{Bourtsoulatze_2019_DJS} and ``ROC-VAE'' \cite{Saidutta_2021_JSC}, which represent traditional autoencoder and variational autoencoder approaches. Note that the training dataset for the ROC-AE, ROC-VAE, \gls{atroc}, and pre-trained \gls{tgcp} is identical. For a fair comparison, ROC-AE, ROC-VAE, and \gls{atroc} use the same network structure, resulting in the same bits per service (i.e., $6144$).
In particular, ROC-VAE and ROC-AE are also fine-tuned by our proposed \gls{jscc} modulation scheme for 64-\gls{qam}, where the optimal constellation parameters $r^*$ are 4 and 50.4, respectively.

\subsection{Results of JSCC Modulation}
\begin{figure*}[htbp]
    \centering
    \subfloat[Training of 16-QAM]{
        \includegraphics[width=0.31\textwidth]{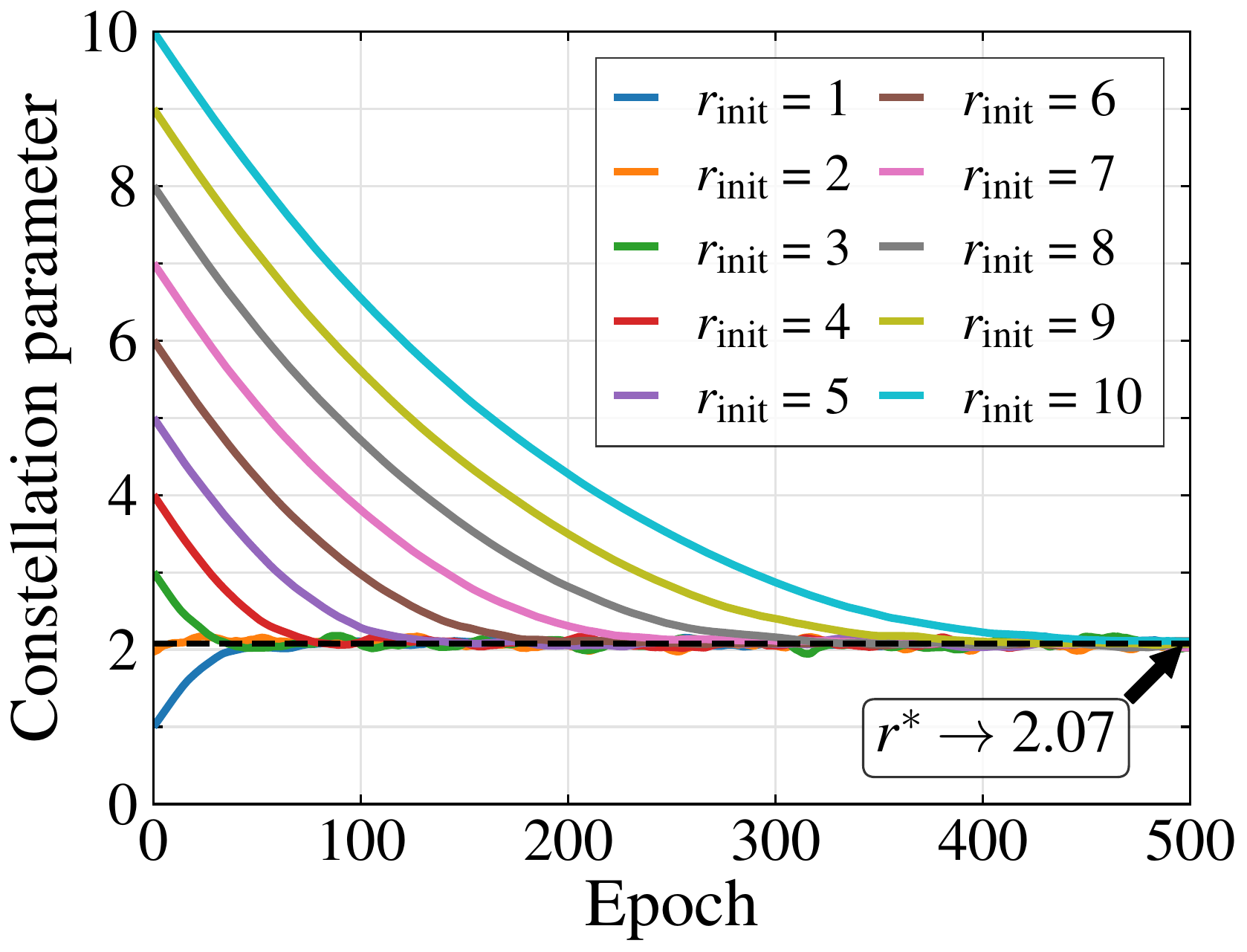}
    }
    \hspace{0.001in}
    \subfloat[Training of 64-QAM.]{
        \includegraphics[width=0.31\textwidth]{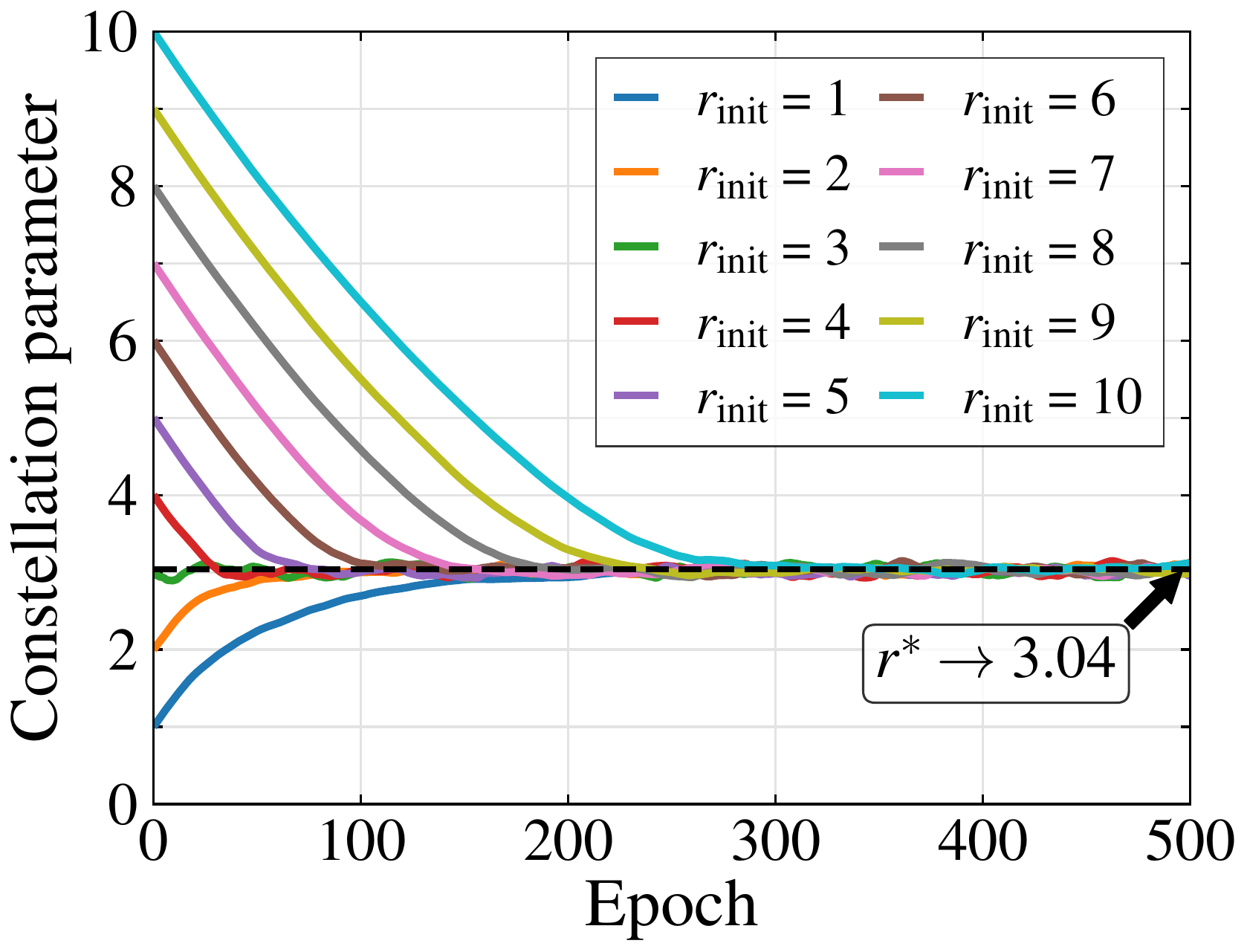}
    }
    \hspace{0.001in}
    \subfloat[Training of 256-QAM.]{
        \includegraphics[width=0.31\textwidth]{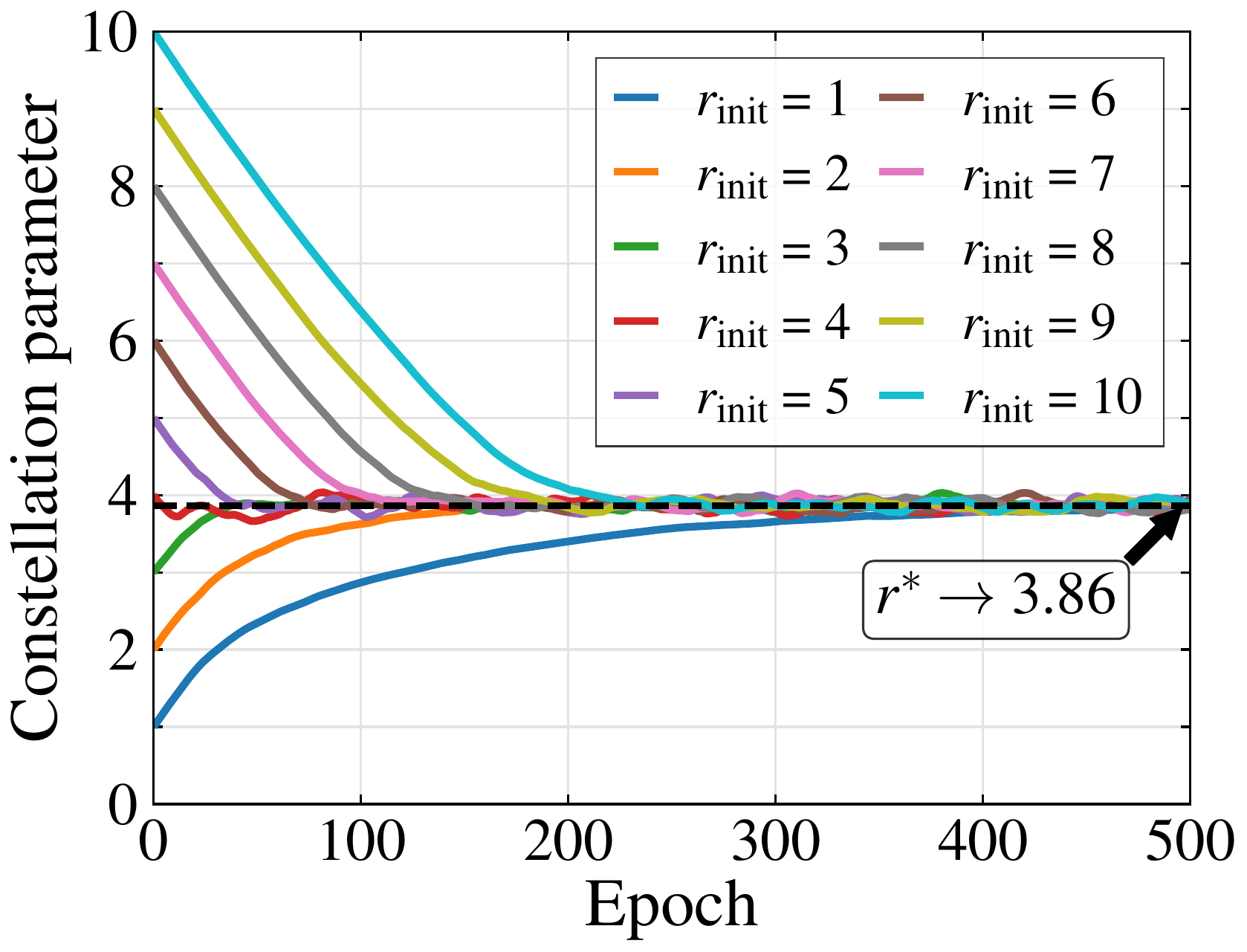}
    }
    \caption{Training of the constellation parameter for 16-QAM, 64-QAM, and 256-QAM. Regardless of the initial value of the constellation parameter, the optimal value consistently converges.}
    \label{fig:Q_train}
\end{figure*}

The constellation parameter $r$ is trained using a pre-trained \gls{jscc} encoder, as described in \cref{alg_quantization}. \Cref{fig:Q_train} demonstrates that regardless of the initial value of the constellation parameter, $r_{\text{init}}\in \{1,\cdots,10 \}$, the optimal constellation parameter $r^*$ consistently converges, validating the effectiveness of the proposed modulation approach.
 
\begin{figure}[t]
\centering
    \begin{center}
    \includegraphics[width=0.36\textwidth]{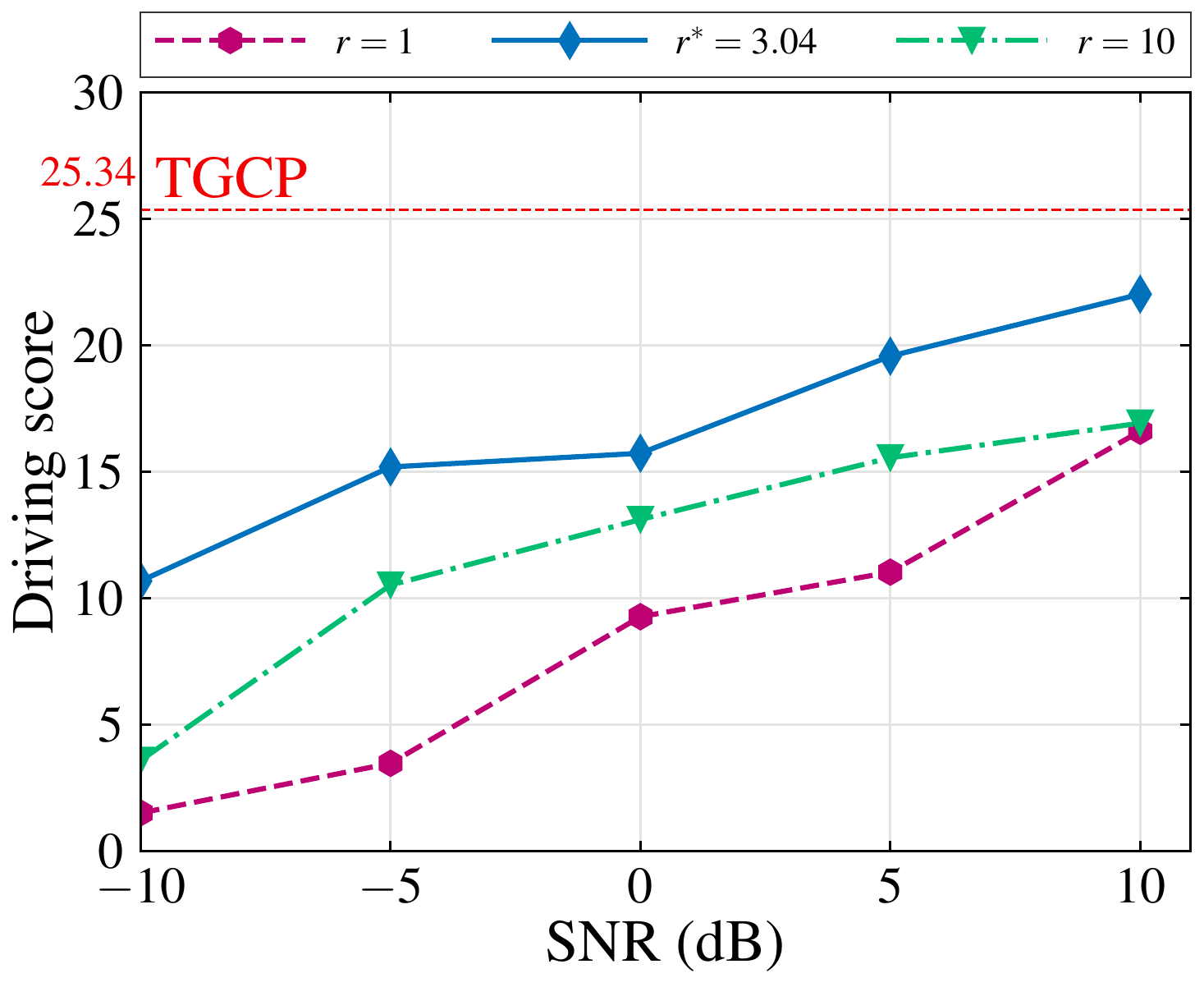}
    \end{center}
       \caption{Driving score of fine-tuned models based on 64-QAM with different constellation parameters ($r\in \{1, r^*, 10\}$, where $r^*=3.04$) under the AWGN channel with SNR range from -10 dB to 10 dB.}
    \label{fig:Q_score}
\end{figure}

Driving scores from different fine-tuned models across various constellation parameters based on 64-\gls{qam} are presented in \cref{fig:Q_score}. The model fine-tuned with the optimal constellation parameter $r^*$ outperforms other models under the \gls{awgn} channel with SNRs range from -10 dB to 10 dB, showcasing the superiority of our proposed \gls{jscc} modulation scheme.

\subsection{Evaluation on CARLA}

\begin{figure}[t]
    \centering
    \subfloat[AWGN channel with SNR = 10 dB.]{
        \includegraphics[width=0.44\textwidth]{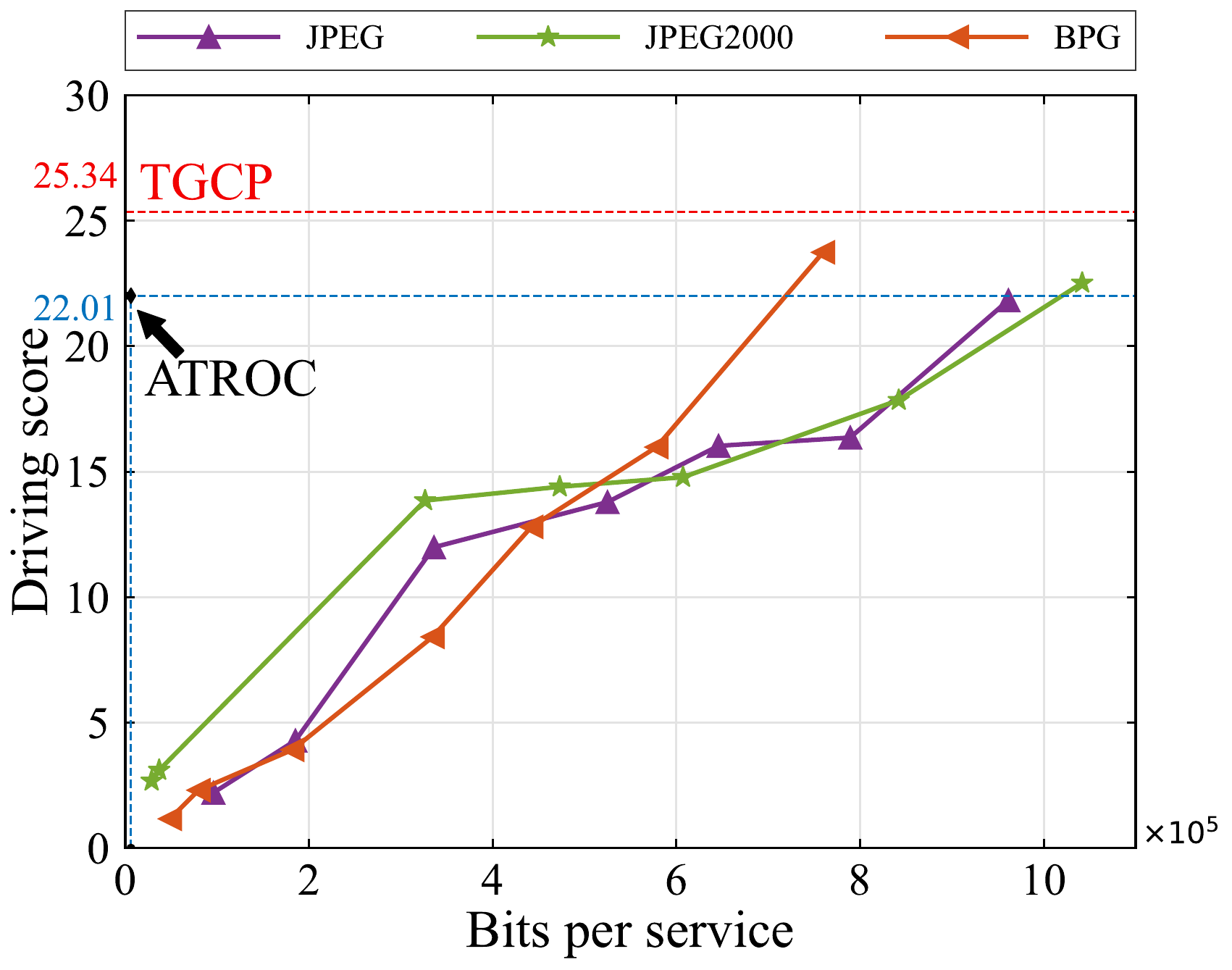}
        \label{subfig:score_bps_AWGN}
    }
    \hfill
    \centering
    \subfloat[Rayleigh fading channel with SNR = 20 dB.]{
        \includegraphics[width=0.44\textwidth]{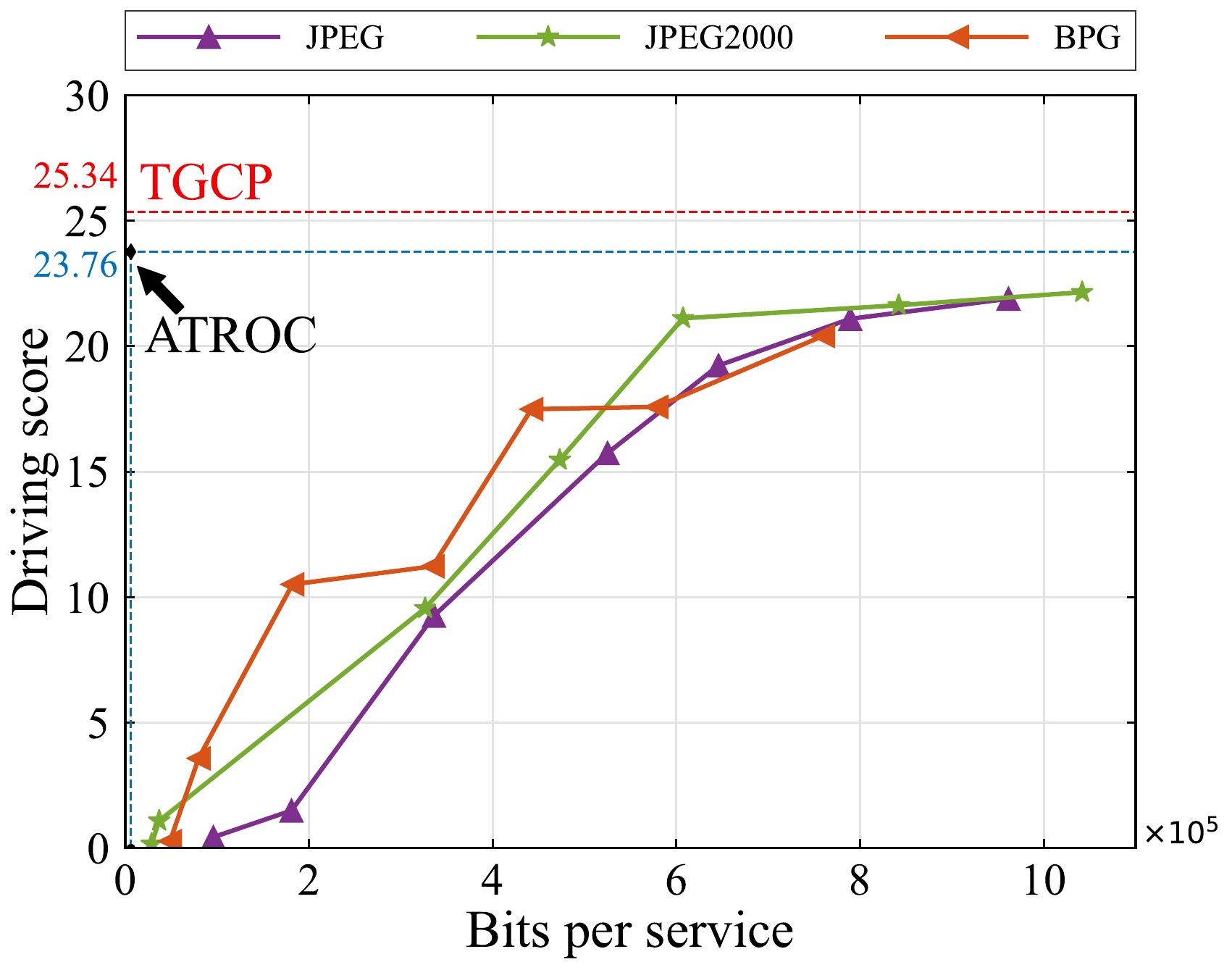}
        \label{subfig:score_bps_Rayleigh}
    }
   \caption{Driving score of traditional reconstruction-oriented communication with varied bits per service under AWGN channel and Rayleigh channel. The ATROC with \(6144\) bits per service serves as a baseline for comparison across both channel conditions. In addition, the TCGP using raw RGB images (\(5.5296\times 10^{6}\) bits per service) for autonomous driving is also included as a baseline.}
    \label{fig:score_bps}
\end{figure}

\begin{figure}[t]
    \centering
    \subfloat[AWGN channel with SNRs range from -10 dB to 10 dB.]{
        \includegraphics[width=0.4\textwidth]{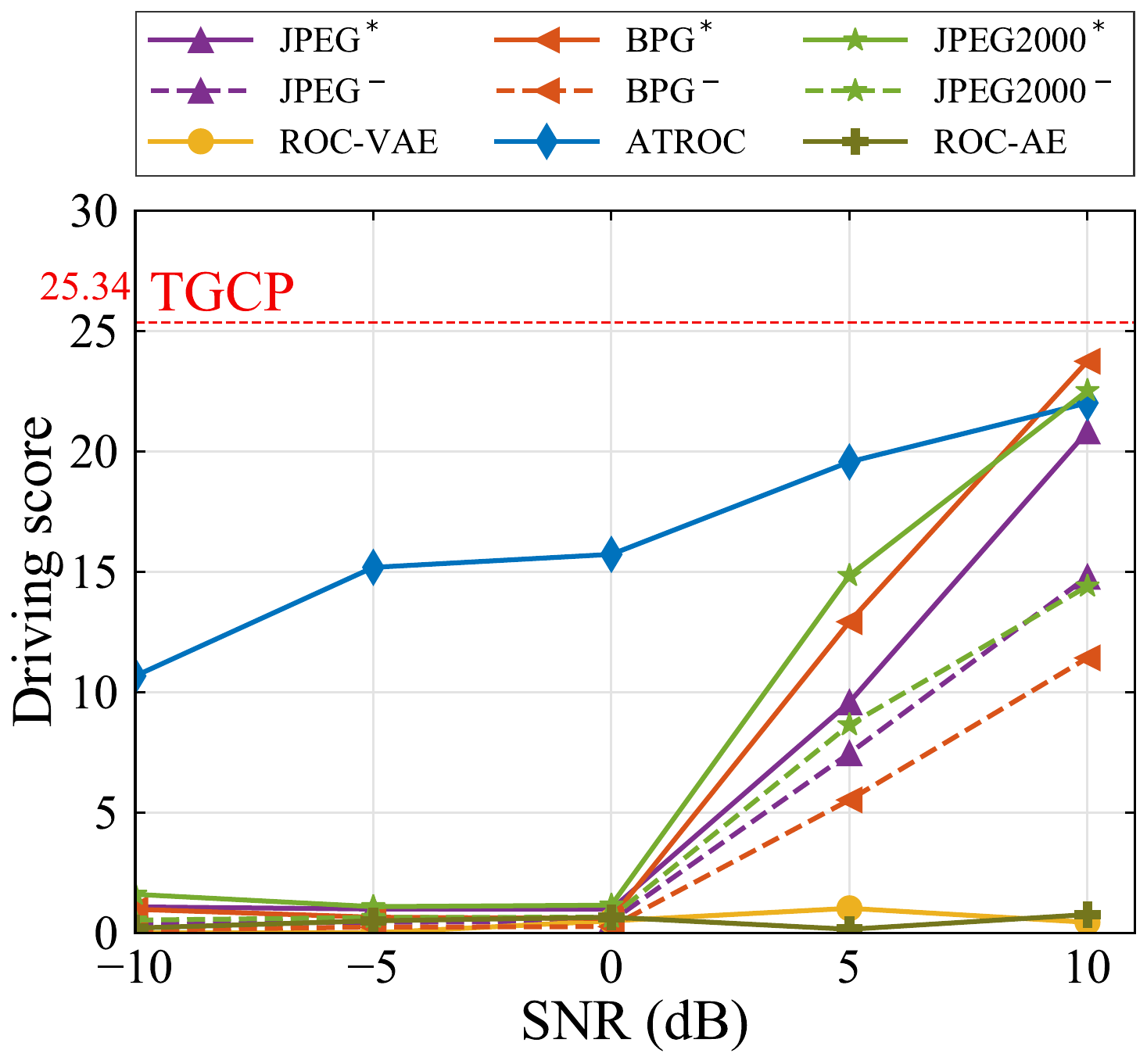}
        \label{fig:score_AWGN}
    }
    \hfill
    \centering
    \subfloat[Rayleigh channel with SNRs range from 0 dB to 20 dB.]{
        \includegraphics[width=0.4\textwidth]{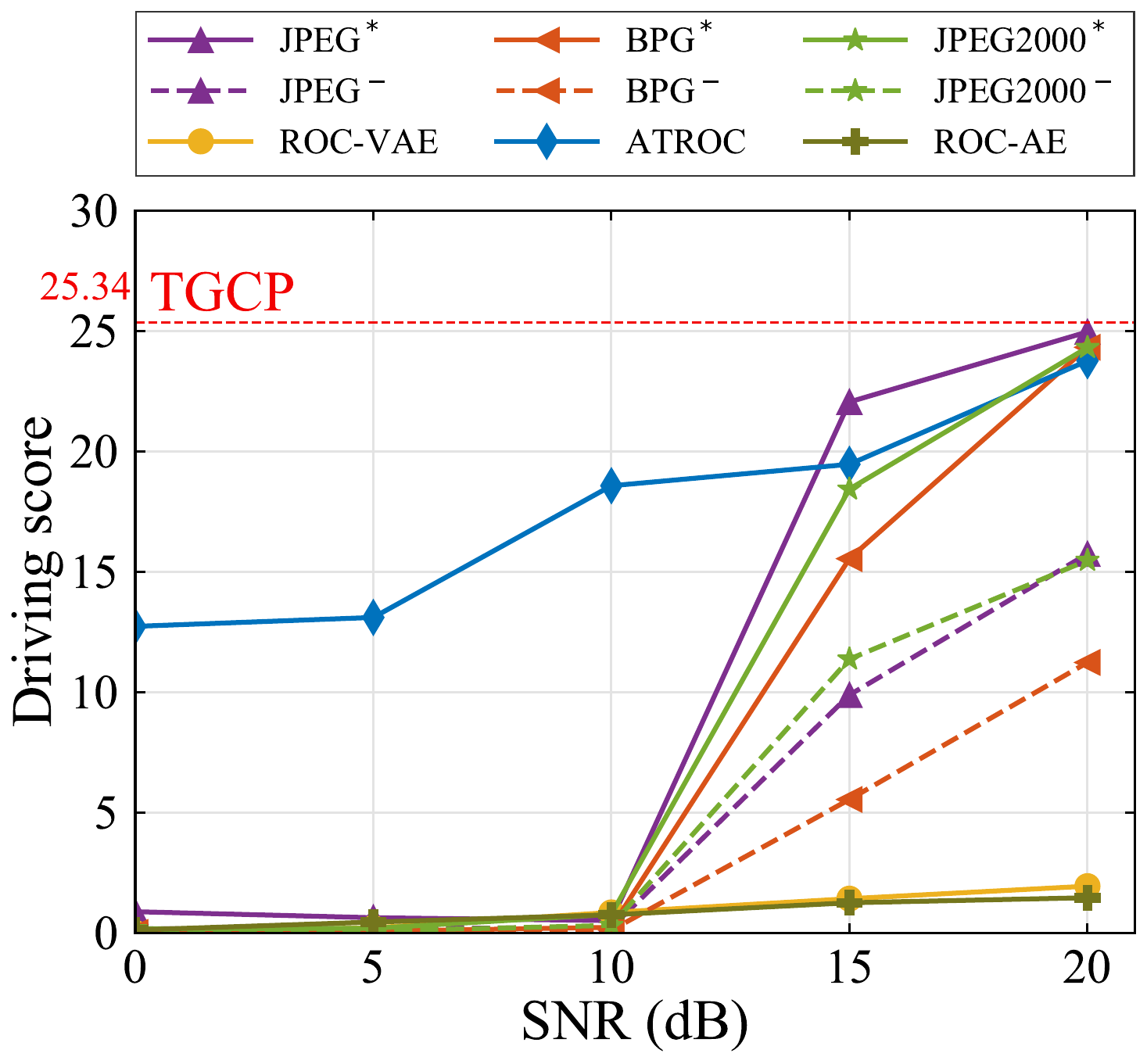}
        \label{fig:score_Rayleigh}
    }
    \caption{Driving score with varied SNRs under AWGN channel and Rayleigh channel.}
\end{figure}

\begin{figure*}[t]
    \centering
    \includegraphics[width=0.8\linewidth]{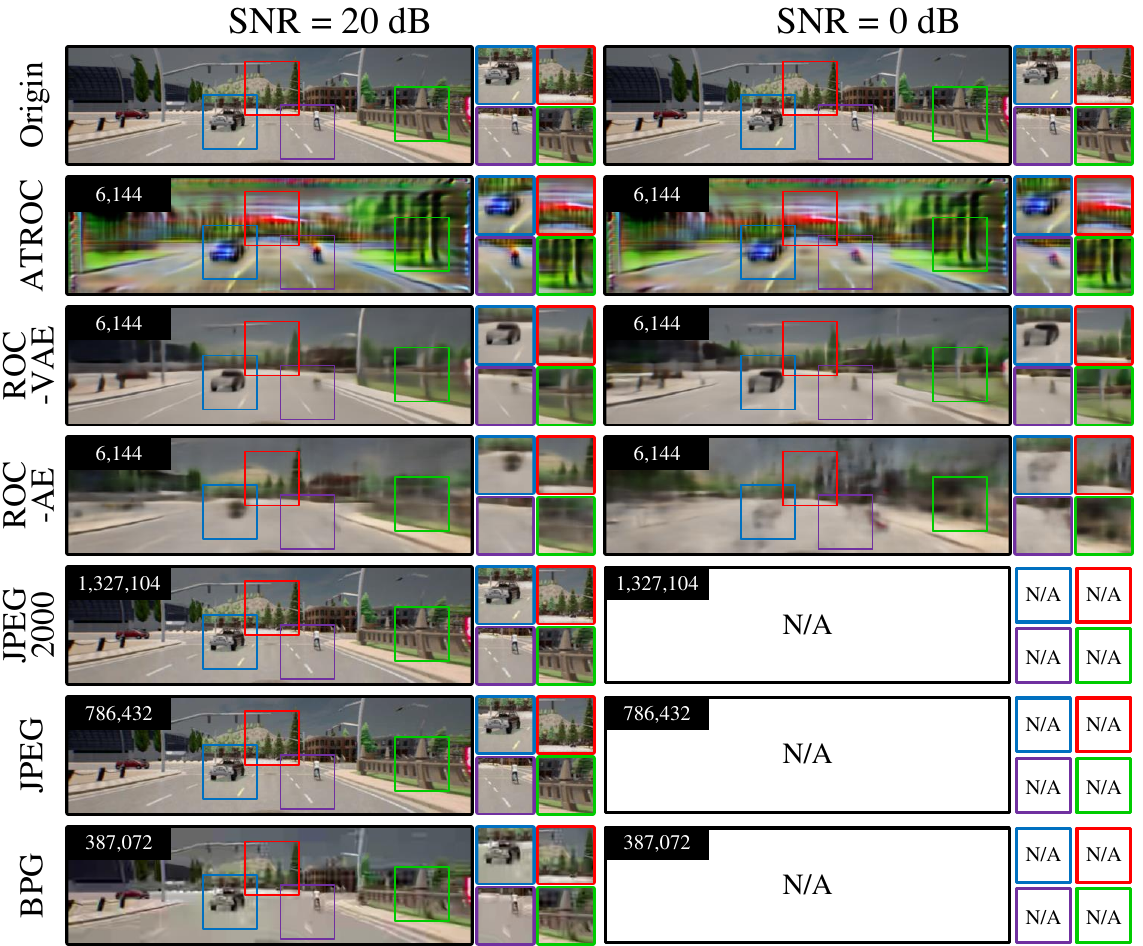}
       \caption{A qualitative example of our proposed method and baseline methods under Rayleigh fading channel with SNR = 20 dB and SNR = 0 dB. The bits per service of each image are provided in the upper left corner. The details in the reconstructed image are highlighted on the right side of the image. 1) blue box: vehicle and road marks; 2) red box: traffic lights; 3) purple box: cyclist and road marks; 4) green box: fence in the distance. Since traditional reconstruction-oriented communication methods (JPEG, JPEG2000, and BPG) fail to reconstruct images when SNR = 0 dB, we use ``N/A'' (Not Applicable) to represent the corrupted images.}
    \label{fig:result_qualitative}
\end{figure*}

The impact of bits per service on the driving score is illustrated in \cref{fig:score_bps}, where the driving score of \gls{tgcp} using raw images without communication is 25.34.
Notably, our proposed method achieves significant bandwidth savings (99.19\% compared to existing methods) while maintaining a required driving score of 20 under both the \gls{awgn} and Rayleigh fading channels. This substantial reduction in bits per service not only illustrates the efficiency of our approach but also underscores its capability to operate effectively under stringent bandwidth constraints. 

Detailed results for specific channel conditions are shown in \cref{fig:score_AWGN} and \cref{fig:score_Rayleigh}, demonstrating the dependency of driving scores on \gls{snr}. For traditional image coding methods, such as JPEG, JPEG2000, and BPG, we apply two configurations for comparison: (1) the average bits per service are set to 961,484 for \(\text{JPEG}^{*}\), 1,041,758 for \(\text{JPEG2000}^{*}\), and 759,683 for \(\text{BPG}^{*}\); (2) the average bits per service are reduced to 524,996 for \(\text{JPEG}^{-}\), 472,958 for \(\text{JPEG2000}^{-}\), and 442,152 for \(\text{BPG}^{-}\). The first configuration highlights the optimal performance of traditional image coding methods, as shown in \cref{subfig:score_bps_AWGN} and \cref{subfig:score_bps_Rayleigh}. In contrast, the second configuration approximately halves the bits per service from the first, as a basis for further comparison.
In contrast, our method requires only 6144 bits per service, highlighting its superior compression and transmission efficiency. In addition, we compare the proposed method \gls{atroc} with the state-of-the-art reconstruction-oriented \gls{jscc} designs using the same neural network structure, with the legends ``ROC-AE'' \cite{Bourtsoulatze_2019_DJS} and ``ROC-VAE'' \cite{Saidutta_2021_JSC}.

Under \gls{awgn} channel conditions, our method significantly outperforms reconstruction-oriented communication methods with driving scores of 15.72 at \gls{snr} = 0 dB, 15.18 at \gls{snr} = -5 dB, and 10.67 at \gls{snr} = -10 dB, as shown in \cref{fig:score_AWGN}. Traditional methods (\(\text{JPEG}^{*}\), \(\text{JPEG}^{-}\),  \(\text{JPEG2000}^{*}\), \(\text{JPEG2000}^{-}\), \(\text{BPG}^{*}\), and \(\text{BPG}^{-}\),) show a dramatic decline in the driving score as \gls{snr} decreases (below 0 dB), emphasizing the robustness of our \gls{atroc} framework under challenging conditions.

Similarly, in Rayleigh fading channel scenarios (\cref{fig:score_Rayleigh}), our proposed method continues to demonstrate superior performance with driving scores of 18.57 at \gls{snr} = 10 dB, 13.1 at \gls{snr} = 5 dB, and 12.73 at \gls{snr} = 0 dB. However, traditional methods experience significant performance degradation when \gls{snr} is below 10 dB.

Moreover, \gls{jscc}-based reconstruction-oriented communication methods perform poorly under this extremely limited communication bandwidth, as these methods fail to preserve task-specific information.

These findings are further supported by qualitative analysis, as illustrated in \cref{fig:result_qualitative}. \gls{jscc}-based reconstruction-oriented communication methods, while capable of producing high-quality image reconstructions suitable for human vision, often fail to retain crucial task-specific information, such as vehicles, cyclists, road markers, and traffic lights. This deficiency leads to poor performance in edge-based autonomous driving applications, where precise detection of such elements is critical for safety and efficiency. In contrast, our proposed method can effectively preserve task-specific information, shown in the blue, red, and purple boxes of \cref{fig:result_qualitative}. To reduce the required bits per service, it ignores task-agnostic information, shown in the green box of \cref{fig:result_qualitative}.
Moreover, our proposed method demonstrates remarkable noise resistance under low \gls{snr} conditions. It effectively preserves task-specific information, maintaining its completeness even in challenging communication environments.

Furthermore, in \Cref{tab_result_metrics}, we evaluate additional performance metrics such as \gls{psnr}, \gls{msssim}, and \gls{fid}, which are typically used to assess image quality from a human perspective. The divergence in performance metrics between traditional reconstruction-oriented methods and our proposed method highlights the necessity of a communication design that prioritizes machine vision, particularly in applications where decision-making accuracy is critical.

\begin{table}
\centering
\footnotesize
\caption{Human Perceptional Metrics}
\label{tab_result_metrics}
\begin{tabular}{ccccc} 
\toprule
Method   & $k\cdot c$ & PSNR(dB) $\uparrow$ & MS-SSIM $\uparrow$ & FID$\downarrow$  \\ 
\midrule
JPEG     & 961484           & 34.56               & 0.99               & 5.83             \\
JPEG2000 & 1041758          & 37.54               & 0.99               & 7.17             \\
BPG      & 759683           & 34.93               & 0.98               & 6.68             \\
ROC-AE   & 6144             & 17.24               & 0.41               & 200.59           \\
ROC-VAE  & 6144             & 21.75               & 0.72               & 135.68           \\
TOC      & 6144             & 11.43               & 0.27               & 268.14           \\
\bottomrule
\end{tabular}
\end{table}

\section{Conclusion}
\label{sec:conclusion}
This paper has investigated an \gls{atroc} framework for edge intelligence, aimed at improving the integration of AI technologies within existing communication infrastructures. By extending the \gls{ib} theory and incorporating \gls{jscc} modulation, our framework shifts the focus from traditional signal reconstruction fidelity to task relevance, thus optimizing the performance of AI-driven applications in bandwidth-constrained and noise-interference environments.

Our evaluations conducted within the CARLA simulator highlight the robustness of the proposed \gls{atroc} framework. Particularly in low \gls{snr} conditions, our framework demonstrated significant superiority over traditional reconstruction-oriented communication methods by achieving a reduction of up to 99.19\% bits per service without sacrificing the effectiveness of task execution. 

The qualitative analysis revealed that while reconstruction-oriented communication methods are effective for human visual perception, they often fail to satisfy the specific requirements of machine vision. This observation emphasizes the need for communication designs that align more closely with the specific information needs of AI systems rather than human interpretation.

\begin{appendices}

\section{Derivation of (\ref{eq_firstTerm_mean})}
\label{apd_fristTerm_mean}
According to the left part of \cref{eq_KL_conditional_2}, we can obtain
\begin{align}
    &\int p(\bm{a}, \bm{y})\log p(\bm{a} | \bm{y}) \dif\bm{a} \dif\bm{y} \notag \\
    =& \int p(\bm{a}, \bm{x}, \bm{z}, \hat{\bm{z}}, \bm{y})\log p(\bm{a} | \bm{y}) \dif\bm{a} \dif\bm{x} \dif\bm{z} \dif\hat{\bm{z}} \dif\bm{y} \notag \\
    =& \int p(\bm{a}, \bm{x}) p(\bm{z}, \hat{\bm{z}}, \bm{y} | \bm{a}, \bm{x})\log p(\bm{a} | \bm{y}) \dif\bm{a} \dif\bm{x} \dif\bm{z} \dif\hat{\bm{z}} \dif\bm{y} \notag
\end{align}
Considering the Markov chain \(A \rightarrow X \rightarrow Z \rightarrow \hat{Z} \rightarrow Y\), we have \( p(\bm{z}, \hat{\bm{z}}, \bm{y} | \bm{a}, \bm{x}) =  p(\bm{z}, \hat{\bm{z}}, \bm{y} | \bm{x})\). Since \(\int p(\bm{z}, \hat{\bm{z}}, \bm{y} | \bm{x}) \dif\bm{z} \dif\hat{\bm{z}} = p(\bm{y}|\bm{x})\), we can obtain

\begin{align}
    & \int p(\bm{a}, \bm{x}) p(\bm{z}, \hat{\bm{z}}, \bm{y} | \bm{a}, \bm{x})\log p(\bm{a} | \bm{y}) \dif\bm{a} \dif\bm{x} \dif\bm{z} \dif\hat{\bm{z}} \dif\bm{y} \notag \\
    =& \int p(\bm{a}, \bm{x}) p(\bm{y} | \bm{x})\log p(\bm{a} | \bm{y}) \dif\bm{a} \dif\bm{x} \dif\bm{y} \notag \\
    =&\ \mathbb{E}_{\bm{a},\bm{x}}\left[\mathbb{E}_{\bm{y}|\bm{x}; \phi,\theta}[\log p(\bm{a} | \bm{y})] \right]. \notag
\end{align}
Similarly, we can obtain
\begin{align}
    \int p(\bm{a},\bm{y})\log q_{\psi}(\bm{a}|\bm{y}) \dif\bm{a}\dif\bm{y} = \mathbb{E}_{\bm{a},\bm{x}}\left[\mathbb{E}_{\bm{y}|\bm{x};\phi,\theta}[\log q_{\psi}(\bm{a}|\bm{y})]\right]. \notag
\end{align}
Based on \cref{eq_KL_conditional_2}, we can obtain
\begin{align}
    \mathbb{E}_{\bm{a},\bm{x}}\big[\mathbb{E}_{\bm{y}|\bm{x}; \phi,\theta}[\log p(\bm{a} | \bm{y})] \big] \geq \mathbb{E}_{\bm{a},\bm{x}}\left[\mathbb{E}_{\bm{y}|\bm{x};\phi,\theta}[\log q_{\psi}(\bm{a}|\bm{y})]\right], \notag
\end{align}
so that
\begin{align}
    \mathbb{E}_{\bm{a},\bm{x}}\big[\mathbb{E}_{\bm{y}|\bm{x}; \phi,\theta}[-&\log p(\bm{a} | \bm{y})] \big] \notag \\
    &\leq \mathbb{E}_{\bm{a},\bm{x}}\left[\mathbb{E}_{\bm{y}|\bm{x};\phi,\theta}[-\log q_{\psi}(\bm{a}|\bm{y})]\right]. \notag
\end{align}

\section{Derivation of (\ref{eq_log2normal})}
\label{apd:derivation_log}
Since we assume $q_{\psi}(\bm{a}|\bm{y})$ takes on a Gaussian distribution $\mathcal{N}(\bm{\mu}_{\psi}(\bm{y}), \bm{\Sigma}_{\psi}(\bm{y}))$, where $\bm{\Sigma}_{\psi}(\bm{y})=\sigma_{c}^{2}I_{d}$, the simplification process is shown below: 
\begin{align}
&-\log{q_{\psi}(\bm{a}|\bm{y})} \notag\\
=& -\log{\mathcal{N}\left(\bm{\mu}_{\psi}(\bm{y}), \bm{\Sigma}_{\psi}(\bm{y})\right)} \notag\\
=& -\log\left[
\exp\left(-{\frac{1}{2}}(\bm{a}-\bm{\mu}_{\psi}(\bm{y}))^{T}\bm{\Sigma}_{\psi}^{-1}(\bm{y})(\bm{a}-\bm{\mu}_{\psi}(\bm{y}))   \right)\right] \notag\\
& + \log\left[(2\pi)^{\frac{d}{2}}|\bm{\Sigma}_{\psi}(\bm{y})|^{\frac{1}{2}} \right] \notag\\
=& \frac{\left\|\bm{a}-\bm{\mu}_{\psi}(\bm{y})\right\|^2_2}{2\sigma^{2}_{c}}  
+ d\log\sigma_{c} + \frac{d}{2}\log{2\pi}. \notag
\end{align}
Since $\sigma_{c}$ is a constant, we have 
\begin{align}
    -\log{q_{\psi}(\bm{a}|\bm{y})}\sim \frac{1}{2\sigma^2_{c}}\|\bm{a}-\bm{\mu}_{\psi}(\bm{y}) \|^2_2. \notag
\end{align}

\end{appendices}

\printbibliography

\end{document}